\documentclass[english,onecolumn]{IEEEtran}
\usepackage[T1]{fontenc}
\usepackage[latin9]{inputenc}
\usepackage{color}
\usepackage{verbatim}
\usepackage{amsthm}
\usepackage{amsmath}
\usepackage{amssymb}
\usepackage{graphicx}
\usepackage{psfrag}
\usepackage{esint}

\makeatletter
\theoremstyle{plain}  
\newtheorem{thm}{\protect\theoremname}
\theoremstyle{definition}
\newtheorem{defn}{\protect\definitionname}
\theoremstyle{definition}
\newtheorem{example}{\protect\examplename}
\theoremstyle{plain}
\newtheorem{prop}{\protect\propositionname}
\theoremstyle{plain}
\newtheorem{lem}{\protect\lemmaname}

\usepackage{xspace}
\usepackage{bbm}
\input{mathlig}

\newcommand{\muspace}{\mspace{1mu}}

\DeclareRobustCommand{\scond}{\mathchoice{\muspace\vert\muspace}{\vert}{\vert}{\vert}}
\mathlig{|}{\scond}

\DeclareRobustCommand{\discint}{\mathchoice{\mspace{-1.5mu}:\mspace{-1.5mu}}{\mspace{-1.5mu}:\mspace{-1.5mu}}{:}{:}}
\mathlig{::}{\discint}

\def\diag{\mathop{\rm diag}\nolimits}%
\newcommand{\xt}{{\tilde{x}}}

\DeclareMathOperator\E{\textsf{E}}
\let\P\relax
\DeclareMathOperator\P{\textsf{P}}

\newcommand{\U}{\mathrm{Unif}}

\def\textiid{i.i.d.\@\xspace}
\newcommand\iid{\ifmmode\text{ i.i.d. } \else \textiid \fi}

\def\mathllap{\mathpalette\mathllapinternal}
\def\mathllapinternal#1#2{%
  \llap{$\mathsurround=0pt#1{#2}$}}

\def\clap#1{\hbox to 0pt{\hss#1\hss}}
\def\mathclap{\mathpalette\mathclapinternal}
\def\mathclapinternal#1#2{%
  \clap{$\mathsurround=0pt#1{#2}$}}

\let\oldstackrel\stackrel
\renewcommand{\stackrel}[2]{\oldstackrel{\mathclap{#1}}{#2}}

\renewcommand{\hbar}{h\mathllap{\overline{\vphantom{h}\hphantom{\rule{4.6pt}{0pt}}}\mspace{0.77mu}}}

\catcode`~=11 
\newcommand{\urltilde}{\kern -.06em\lower -.06em\hbox{~}\kern .02em}
\catcode`~=13 

\usepackage{babel}
\providecommand{\definitionname}{Definition}
\providecommand{\examplename}{Example}
\providecommand{\lemmaname}{Lemma}
\providecommand{\propositionname}{Proposition}
\providecommand{\theoremname}{Theorem}

\providecommand{\examplename}{Example}

\usepackage{babel}
\providecommand{\definitionname}{Definition}
\providecommand{\examplename}{Example}
\providecommand{\lemmaname}{Lemma}
\providecommand{\propositionname}{Proposition}
\providecommand{\theoremname}{Theorem}

\usepackage{babel}
\providecommand{\definitionname}{Definition}
\providecommand{\examplename}{Example}
\providecommand{\lemmaname}{Lemma}
\providecommand{\propositionname}{Proposition}
\providecommand{\theoremname}{Theorem}

\makeatother

\usepackage{babel}
\providecommand{\definitionname}{Definition}
\providecommand{\examplename}{Example}
\providecommand{\lemmaname}{Lemma}
\providecommand{\propositionname}{Proposition}
\providecommand{\theoremname}{Theorem}

\begin{document}

\title{Distributed Simulation of Continuous Random Variables}

\author{Cheuk Ting Li and Abbas El Gamal\\
 Department of Electrical Engineering, 
 Stanford University\\
 Email: ctli@stanford.edu, abbas@ee.stanford.edu 
   \thanks{
    The work of C. T. Li was partially supported by a Hong Kong Alumni Stanford Graduate Fellowship. This paper was presented in part at the IEEE International Symposium on Information Theory, Barcelona, July 2016.}
   }

%

\maketitle
\begin{abstract}
We establish the first known upper bound on the exact and Wyner's
common information of $n$ continuous random variables
in terms of the dual total correlation between them (which is a generalization
of mutual information). In particular, we show that when the pdf of
the random variables is log-concave, there is a constant gap of $n^{2}\log e+9n\log n$
between this upper bound and the dual total correlation lower bound
that does not depend on the distribution. The upper bound is obtained
using a computationally efficient dyadic decomposition scheme for
constructing a discrete common randomness variable $W$ from which
the $n$ random variables can be simulated in a distributed manner.
We then bound the entropy of $W$ using a new measure, which we refer
to as the erosion entropy.
\end{abstract}


\section{Introduction}

This paper is motivated by the following question. Alice would like
to simulate a random variable $X_{1}$ and Bob would like to simulate
another random variable $X_{2}$ such that $(X_{1},X_{2})$ are jointly
Gaussian with a prescribed mean and covariance matrix. Can these two
random variables be simulated in a distributed manner with only a
finite amount of common randomness between Alice and Bob?

We answer this question in the affirmative for $n$
continuous random variables under certain conditions on their joint pdf, including when it is log-concave such as Gaussian.

\begin{figure}[h!]
\begin{center}
\small 
\psfrag{w}[l]{$W\in \{0,1\}^*$} 
\psfrag{e}[c]{Common Randomness} 
\psfrag{r}[c]{Source}
\psfrag{m}[l]{$W$} 
\psfrag{d1}[c]{Agent 1} 
\psfrag{d2}[c]{Agent 2} 
\psfrag{dn}[c]{Agent $n$} 
\psfrag{x1}[t]{$X_{1}$} 
\psfrag{x2}[t]{$X_{2}$}
\psfrag{xn}[t]{$X_{n}$}
\includegraphics[scale=0.53]{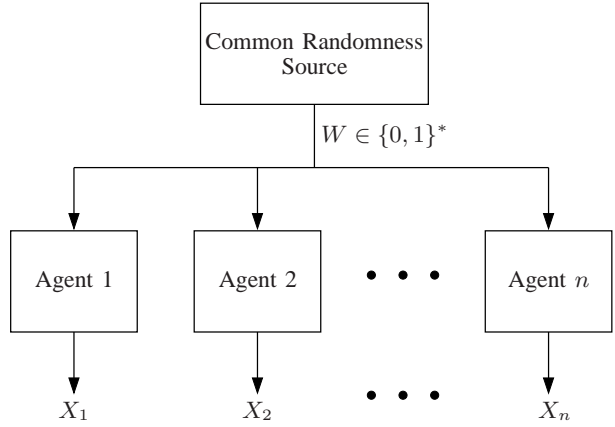}
\caption{Distributed randomness generation setting. Common randomness
$W$ is broadcast to $n$ agents and agent $i\in[1:n]$ generates
$X_{i}$ using $W$ and it local randomness.}
\label{fig:setting} 
\end{center}
\end{figure}

The general distributed randomness generation setup we consider is
depicted in Figure~\ref{fig:setting}. There are $n$ agents (e.g.,
processors in a computer cluster or nodes in a communication network)
that have access to \emph{common randomness} $W\in\{0,1\}^{*}$. Agent
$i\in[1:n]$ wishes to simulate the random variable $X_{i}$ using
$W$ and its local randomness, which is independent of $W$ and the local
randomness at other agents, such that $X^n=(X_{1},\ldots,X_{n})$
follows a prescribed distribution \emph{exactly}. The distributed
randomness simulation problem is to find the common randomness $W^{*}$
with the minimum average description length $R^{*}$, referred to
in~\cite{kumar2014exact} as the \emph{exact common information}
between $X^n$, and the scheme that achieves
this exact common information.

Since $W$ can be represented by an optimal prefix-free code, e.g.,
a Huffman code or the code in~\cite{linder1997existence} if the alphabet
is infinite, the average description length $R^{*}$ can be upper bounded
as $H(\tilde{W})\le R^{*}<H(\tilde{W})+1$, where $\tilde{W}$ minimizes
$H(W)$. Hence in this paper we will focus on investigating $W$ that
minimizes $H(W)$ instead of $R^{*}$.

%

The above setting was introduced in~\cite{kumar2014exact} for two
discrete random variables and the minimum entropy of $W$, referred
to as the \emph{common entropy}, is given by 
\begin{equation}
G(X_{1};X_{2})=\min_{W:\,X_{1}\perp\!\!\!\perp X_{2}|W}H(W).\label{eqn:g2}
\end{equation}
A similar quantity for channel simulation was also studied by Harsha et. al.~\cite{harsha2007communication}.
Computing $G(X_{1};X_{2})$, even for moderate size random variable
alphabets, can be computationally difficult since it involves minimizing
a concave function over a non-convex set; see~\cite{kumar2014exact}
for some cases where $G$ can be computed and for some properties
that can be exploited to compute it. Hence the main difficulty in
constructing a scheme that achieves $G$ (within 1-bit) for a given
$(X_{1},X_{2})$ distribution is finding the optimal common randomness
$W$ that achieves it.

It can be readily shown that
\begin{equation}
I(X_{1};X_{2})\le J(X_{1};X_{2})\le G(X_{1};X_{2})\le\min\{H(X_{1}),H(X_{2})\},\label{eqn:bounds2}
\end{equation}
where 
\begin{equation}
J(X_{1};X_{2})=\min_{W:\,X_{1}\perp\!\!\!\perp X_{2}|W}I(W;X_{1},X_{2})\label{eqn:j2}
\end{equation}
is Wyner's common information~\cite{Wyner1975a}, which is the
minimum amount of common randomness rate needed to generate the discrete
memoryless source (DMS) $(X_{1},X_{2})$ with asymptotically vanishing
total variation. The notion of exact common information rate $\bar{G}(X_{1};X_{2})$,
which is the minimum amount of common randomness rate needed to generate
the DMS $(X_{1},X_{2})$ exactly, was
also introduced in~\cite{kumar2014exact}. It was shown that: (i) in general $J\le\bar{G}\le G$, (ii) $G$ can be strictly larger than $\bar{G}$, and (iii) in some
cases $\bar{G}(X_{1};X_{2})=J(X_{1};X_{2})$. It is not known, however,
if $\bar{G}(X_{1};X_{2})=J(X_{1};X_{2})$ in general. As such, we
do not consider $\bar{G}$ further in this paper.

The above results can be extended to $n$ random variables. First,
it is straightforward to extend the common entropy in~\eqref{eqn:g2} to $n$ general random
variables to obtain 
\begin{equation}
G(X_{1};\ldots;X_{n})=\min_{W:\,X_{1}\perp\!\!\!\perp X_{2}\perp\!\!\!\perp\ldots\perp\!\!\!\perp X_{n}|W}H(W).\label{eqn:gn}
\end{equation}
Second, in~\cite{cuff2010coordination,liu2010common} Wyner's common information was
extended to $n$ discrete random variables to obtain 
\[
J(X_{1};\ldots;X_{n})=\min_{W:\,X_{1}\perp\!\!\!\perp X_{2}\perp\!\!\!\perp\ldots\perp\!\!\!\perp X_{n}|W}I(W;X_{1},\ldots,X_{n}).
\]
The operational implications of Wyner's common information for two
continuous random variables was studied in~\cite{xu2011wyners}.
Wyner's common information between scalar jointly Gaussian random variables is computed in~\cite{xu2011wyners},
and the result is extended to Gaussian vectors in~\cite{satpathy2015gaussian}, and to outputs of additive Gaussian channels in~\cite{yang2014wyner}.

We can also generalize the bounds in \eqref{eqn:bounds2} to $n$
random variables to obtain 
\begin{equation}
I_{\mathrm{D}}(X_{1};X_{2};\ldots;X_{n})\le J(X_{1};X_{2};\ldots;X_{n})\le G(X_{1};X_{2};\ldots;X_{n})\le\min_{i}H(X_{1},\ldots,X_{i-1},X_{i+1},\ldots X_{n}),\label{eqn:boundsn}
\end{equation}
where $I_{\mathrm{D}}$ is the \emph{dual total correlation}~\cite{te1978nonnegative}\,---\,a
generalization of mutual information defined as
\[
I_{\mathrm{D}}(X_{1};X_{2};\ldots;X_{n})=H(X_{1},\ldots,X_{n})-\sum_{i=1}^{n}H(X_{i}|X_{1},\ldots,X_{i-1},X_{i+1},\ldots,X_{n}).
\]
Details of the derivation of the lower bound in~\eqref{eqn:boundsn}
can be found in Appendix~\ref{sub:jd_ci}. Note that the lower bound
on $J$ continues to hold for continuous random variables after replacing
the entropy $H$ in the definition of $I_{\mathrm{D}}$ with the differential
entropy $h$. There is no corresponding upper bound to \eqref{eqn:boundsn}
for continuous random variables, however, and it is unclear under
what conditions $G$ is finite.

In this paper we devise a computationally efficient scheme for constructing
a common randomness variable $W$ for distributed simulation of $n$
continuous random variables and establish upper bounds on its entropy, which in turn provide upper bounds on $G$.
In particular we establish
the following upper bound on $G$ when the pdf of $X^n$
is \emph{log-concave} 
\[
I_{\mathrm{D}}\le J\le G\le I_{\mathrm{D}}+n^{2}\log e+9n\log n.
\]
For $n=2$, this bound reduces to 
\[
I(X_{1};X_{2})\le J(X_{1};X_{2})\le G(X_{1};X_{2})\le I(X_{1};X_{2})+24.
\]
Applying this result to two jointly Gaussian random variables shows
that only a finite amount of common randomness is needed for their
distributed simulation. The above upper bound also provides an upper
bound on Wyner's common information between $n$ continuous random
variables with log-concave pdf. This is an interesting result since
computing Wyner's common information for $n$ continuous random variables
is very difficult in general and there is no previously known upper
bound on it.

Our distributed randomness simulation scheme uses a dyadic decomposition
procedure to construct the common randomness variable $W$. For $X^n$ uniformly distributed
over a set $A$, our decomposition method partitions $A$ into hypercubes. The common randomness $W$ is defined as the position and size of the hypercube that contains $X_{1},\ldots,X_{n}$ represented via an optimal prefix-free code. 
Conditioned on $W$, the random variables
$X^n$ are independent and uniformly distributed over line segments, which when combined with local randomness facilitates distributed exact simulation. This scheme is extended to non-uniformly distributed
$X^n$ by performing the same dyadic decomposition on the positive part of the hypograph of the pdf of $X^n$. Since bounding $H(W)$ directly is quite difficult, we bound it using the \emph{erosion entropy} of the set, which is a new measure that is shift invariant.

The cardinality of the random variable $W$ needed
for exact distributed simulation of continuous random variables is
in general infinite. By terminating the dyadic
decomposition at a finite iteration, however, we show that the random variables
can be approximately simulated using a fixed length
code such that for log-concave pdfs, the total variation distance between
the simulated and prescribed pdfs can be bounded as a function
of the dual total correlation and the cardinality of $W$. This result provides an upper bound on the one-shot version of Wyner's common information with total variation constraint.

The rest of the paper is organized as follows. In Section \ref{sec:uniform},
we introduce the aforementioned dyadic decomposition scheme and establish
an upper bound on $G$ when the random variables are uniformly distributed
over an orthogonally convex set. In Section \ref{sec:logconcave},
we extend this bound to non-uniform distributions with orthogonally
concave pdf and establish our main result on log-concave pdfs. In
Section \ref{sec:approx_bdd}, we establish an upper bound on the
one-shot version of Wyner's common information with total variation constraint. In Appendix \ref{sec:schemes},
we provide details on the implementation of the coding scheme for
constructing the common randomness variable.

\medskip{}


\subsection{Notation}

Throughout this paper, we assume that $\log$ is base 2 and the entropy
$H$ is in bits. We use the notation: 
$[a:b]=[a,b]\cap\mathbb{Z}$ and $X_{[1:n]\backslash i}=(X_{1},\ldots,X_{i-1},X_{i+1},\ldots,X_{n})$.

A set $A\subseteq\mathbb{R}^{n}$ is said to be \emph{orthogonally
convex} if for any line $L$ parallel to one of the $n$ axes, $L\cap A$
is a connected set (empty, a point, or an interval). A function $f$
is said to be orthogonally concave if its \emph{hypograph} $\left\{ (x,\alpha):\, x\in\mathbb{R}^{n},\,\alpha\le f(x)\right\} $
is orthogonally convex.

We denote the $i$-th standard basis vector of $\mathbb{R}^{n}$ by
$\mathrm{e}_{i}$. We denote the volume of a Lebesgue measurable set $A\subseteq\mathbb{R}^{n}$
 by $\mathrm{V}_{n}(A)=\int_{\mathbb{R}^{n}}\mathbf{1}_{A}(x)dx$.
If $A\subseteq B\subseteq\mathbb{R}^{n}$, where $B$ is an $m$-dimensional
affine subspace, we denote the $m$-dimensional volume of $A$ by
$\mathrm{V}_{m}(A)=\int_{B}\mathbf{1}_{A}(x)dx$.

We define the projection of a point $x\in\mathbb{R}^{n}$ as 
\[
\mathrm{P}_{i_{1},\ldots,i_{k}}(x)=(x_{i_{1}},\ldots,x_{i_{k}})\in\mathbb{R}^{k},
\]
and the projection of a set $A\subseteq\mathbb{R}^{n}$
onto the dimensions $i_{1},\ldots,i_{k}$ as 
\[
\mathrm{P}_{i_{1},\ldots,i_{k}}(A)=\left\{ (x_{i_{1}},\ldots,x_{i_{k}})\,:\, x\in A\right\} \subseteq\mathbb{R}^{k}.
\]
We use the shorthand notation 
\begin{align*}
\mathrm{P}_{\backslash i}(A) & =\mathrm{P}_{1,2,\ldots,i-1,i+1,\ldots,n}(A),\\
\mathrm{VP}_{i_{1},\ldots,i_{k}}(A) & =\mathrm{V}_{k}(\mathrm{P}_{i_{1},\ldots,i_{k}}(A)),\\
\mathrm{VP}_{\backslash i}(A) & =\mathrm{V}_{n-1}(\mathrm{P}_{\backslash i}(A)).
\end{align*}
For $A,B\subseteq\mathbb{R}^{n}$, $A+B$ denotes the Minkowski sum
$\left\{ a+b:\, a\in A,\, b\in B\right\} $, and for $x\in\mathbb{R}^{n}$,
$A+x=\left\{ a+x:\, a\in A\right\} $. For $\gamma\in\mathbb{R}$,
$\gamma A=\left\{ \gamma a:\, a\in A\right\} $. For $M\in\mathbb{R}^{n\times n}$,
$MA=\left\{ Ma:\, a\in A\right\} $. The \emph{erosion} of the set
$A$ by the set $B$ is defined as $A\ominus B=\left\{ x\in\mathbb{R}^{n}:\, B+x\subseteq A\right\} $.

For a set $A\subseteq\mathbb{R}^{n}$ where $0\in A$, the radial
function $\rho_{A}:\mathbb{R}^{n}\to\mathbb{R}$ is defined as $\rho_{A}(x)=\sup\left\{ \lambda\ge0\,:\,\lambda x\in A\right\} $.

\section{Uniform Distribution over a Set \label{sec:uniform}}

We first define the dyadic decomposition of a set, which is the building
block of our distributed randomness simulation scheme. 
\begin{defn}[Dyadic decomposition]
For $v\in\mathbb{Z}^{n}$ and $k\in\mathbb{Z}$, we define the hypercube
$C_{k,v}=2^{-k}([0,1]^{n}+v)\subset\mathbb{R}^{n}$. For a set
$A\subseteq\mathbb{R}^{n}$ with a boundary of measure zero and $k\in\mathbb{Z}$, define the set
\[
D_{k}(A)=\left\{ v\in\mathbb{Z}^{n}\,:\, C_{k,v}\subseteq A\text{ and }C_{k-1,\left\lfloor v/2\right\rfloor }\nsubseteq A\right\} ,
\]
where $\left\lfloor v/2\right\rfloor $ is the vector formed by the
entries $\left\lfloor v_{i}/2\right\rfloor $. 

The \emph{dyadic decomposition}
of $A$ is the partitioning of $A$ into hypercubes $\left\{ C_{k,v}\right\}$
such that $v\in D_{k}(A)$ and $k\in\mathbb{Z}$.
Since every point $x$ in the interior of $A$ is contained in some hypercube in $A$, the interior is contained in 
 $\cup_{k\in\mathbb{Z},\,v\in D_{k}(A)}C_{k,v}$, and the set of points in $A$ not covered by the hypercubes has measure zero.
\end{defn}
For $X^{n}\sim \U(A)$, denote by $C_{K,V}$, $V\in D_{K}(A)$,
the hypercube that contains $X^{n}$ and let the \emph{dyadic decomposition
random variable} $W_{A}=(K,V)$. Then conditioned on $W_{A}=(k,v)$,
$X^{n}\sim \U(C_{k,v})$, that is, $X_{1},\ldots,X_{n}$
are conditionally independent given $W_{A}$. Hence, we can use the
dyadic decomposition to perform distributed randomness simulation as follows. 
\begin{enumerate}
\item The common randomness source generates $\xt^{n}$ according to a uniform
pdf over $A$ and finds $w_{A}=(k,v)$ such that $v\in D_{k}(A)$
and $\xt^{n}\in C_{k,v}$. 
\item The common randomness source represents $w_{A}$ by a codeword from an agreed upon optimal prefix-free code
and sends it to the processors (from this point on, we will assume
that $W_{A}$ is always represented by an optimal prefix-free code). 
\item Upon receiving and recovering $w_{A}=(k,v)$, agent $i$ generates
$X_{i}\sim\U[2^{-k}v_{i},\,2^{-k}(v_{i}+1)]$. 
\end{enumerate}
The implementation details of this scheme are provided in Appendix~\ref{sec:schemes}.

\begin{figure}[h]
\centering{}\includegraphics[scale=0.45]{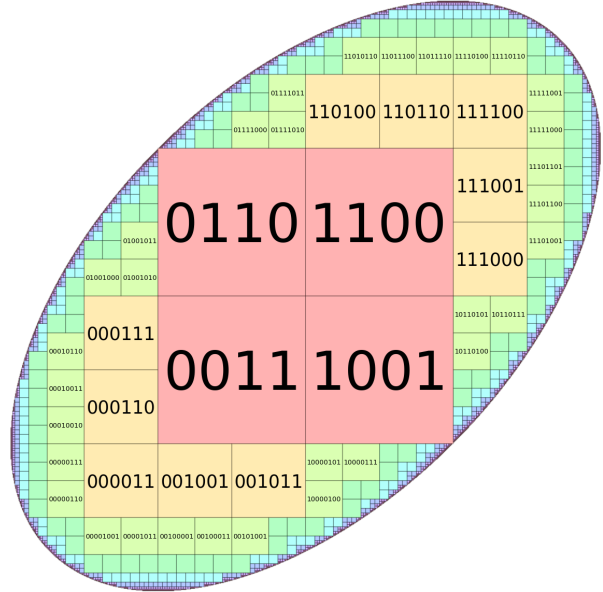} \caption{\label{fig:decomp_ellipse}Dyadic decomposition of the uniform pdf over the ellipse in
Example~\ref{exa:ellipse1}.}
\end{figure}

\begin{figure}[h]
\centering{}\includegraphics[scale=0.4]{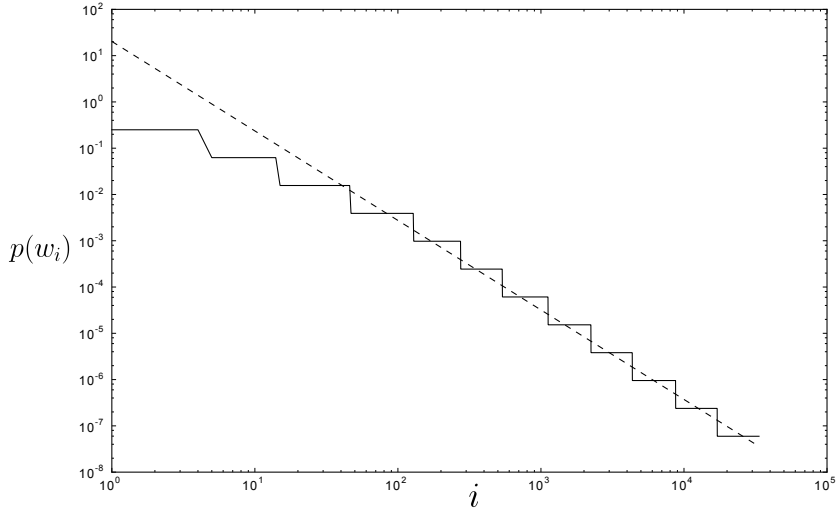} \caption{\label{fig:decomp_ellipse_bar}The pmf of $W$ for the dyadic decomposition
of the uniform pdf over the ellipse in Example~\ref{exa:ellipse1}.}
\end{figure}

To illustrate the above dyadic decomposition scheme, consider the
following. 
\begin{example}
\label{exa:ellipse1} Let $X^{n}\sim\U(A)$, where $A$
is an ellipse, i.e., $A=\left\{ x\in\mathbb{R}^{2}:\, x^{T}Kx<1\right\} $ and
$K$ is a positive definite matrix. Figure~\ref{fig:decomp_ellipse}
illustrates the dyadic decomposition for $K=\left[\begin{array}{cc}
4/3 & -2/3\\
-2/3 & 4/3
\end{array}\right]$, and the codewords assigned to the larger squares. 
Figure~\ref{fig:decomp_ellipse_bar} plots the pmf of the constructed
$W_{A}$ for the same $K$ in a log-log scale ($w_{i}$ is the $i$-th most probable $w$). As can be seen, the tail
of the pmf of $W_{A}$ roughly follows a straight line, that is, the pmf of $W_{A}$ follows a power law tail $p(w_{i})\propto i^{-\alpha}$ with $\alpha\approx2$). Hence $H(W_{A})$ is finite. 

\end{example}

The entropy of $W_{A}$ can be expressed as
\begin{align*}
H(W_{A}) & =-\sum_{k\in\mathbb{Z}}\sum_{v\in D_{k}(A)}\P\left\{ W_{A}=(k,v)\right\} \log\P\left\{ W_{A}=(k,v)\right\} \\
 & =-\sum_{k\in\mathbb{Z}}\sum_{v\in D_{k}(A)}\frac{2^{-nk}}{\mathrm{V}_{n}(A)}\log\frac{2^{-nk}}{\mathrm{V}_{n}(A)}\\
 & =\sum_{k\in\mathbb{Z}}\frac{2^{-nk}\left|D_{k}(A)\right|}{\mathrm{V}_{n}(A)}\left(nk+\log\mathrm{V}_{n}(A)\right)\\
 & =\log\mathrm{V}_{n}(A)+\frac{1}{\mathrm{V}_{n}(A)}\sum_{k\in\mathbb{Z}}nk2^{-nk}\left|D_{k}(A)\right|.
\end{align*}
Since $X_1,\ldots,X_n$ are conditionally independent given $W_{A}$, we have
\[
G(X_{1};\ldots;X_{n})\le H(W_{A}),
\]
and since an optimal prefix-free code is used to represent the  hypercubes resulting from the dyadic
decomposition, the average code length is upper-bounded
by $H(W_{A})+1$.

\medskip{}

The exact value of $H(W_{A})$ is very difficult to compute in general.
It also varies as we shift and scale $A$, which should not matter
in the context of distributed randomness simulation, since we can simply
shift and scale the random variables before applying the scheme. Hence,
we bound $H(W_A)$ using the following quantity, which is shift invariant and easier to analyze.

\begin{defn}[Erosion entropy]
The \emph{erosion entropy}
of a set $A\subseteq\mathbb{R}^{n}$ with $0<\mathrm{V}_{n}(A)<\infty$
by a convex set $B\subseteq\mathbb{R}^{n}$ is defined as
\[
h_{\ominus B}(A)=\int_{-\infty}^{\infty}\left(\mathbf{1}\left\{ t\ge0\right\} -\frac{\mathrm{V}_{n}\left(A\ominus2^{-t}B\right)}{\mathrm{V}_{n}(A)}\right)dt,
\]
where $A\ominus B=\left\{ x\in\mathbb{R}^{n}:\, B+x\subseteq A\right\} $
is the erosion of $A$ by $B$. 
\end{defn}
The erosion entropy roughly measures the ratio of the surface area to the volume of the set $A$. To see this, assume that $0\in B$ and let $X^n\sim \U(A)$ and $\Phi=\sup\{\phi\,:\,\phi B+X\subseteq A\}$, then we can rewrite the erosion entropy as $h_{\ominus B}(A)=\E(-\log(\Phi))$. If we further assume that $B=[0,1]^n$, then the erosion entropy is the expectation of the negative logarithm of the side length of the largest hypercube centered at a randomly distributed point in $A$. Hence, a large erosion entropy means that the largest hypercube centered at a random point in $A$ is small, which suggests that $A$ has a large surface area to volume ratio.
We now state some basic properties of the erosion entropy. 
\begin{prop}
\label{prop:erosion_prop}For a set $A\subseteq\mathbb{R}^{n}$
with $0<\mathrm{V}_{n}(A)<\infty$, convex sets $B,B_{1},B_{2}\subseteq\mathbb{R}^{n}$, and nonsingular $M\in \mathbb{R}^{n \times n}$,
the erosion entropy satisfies the following. 
\begin{enumerate}
\item Monotonicity. If $B_{1}\subseteq B_{2}$, then $h_{\ominus B_{1}}(A)\le h_{\ominus B_{2}}(A)$. 
\item Scaling. $h_{\ominus\beta B}(\alpha A)=h_{\ominus B}(A)+\log(\beta/\alpha)$. 
\item Linear transformation. $h_{\ominus MB}(MA)=h_{\ominus B}(A)$. 
\item Union. If $A_{1},\ldots,A_{k}\subseteq\mathbb{R}^{n}$ are disjoint,
then 
\[
h_{\ominus B}\left(\bigcup_{i=1}^{k}A_{i}\right)\le\sum_{i=1}^{k}\frac{\mathrm{V}_{n}(A_{i})}{\mathrm{V}_{n}(\bigcup_{j}A_{j})}\cdot h_{\ominus B}(A_{i}).
\]
Equality holds when the closures of $A_{1},\ldots,A_{k}\subseteq\mathbb{R}^{n}$
are disjoint. 
\item Reduction to differential entropy. If $X^{n}\sim\U(A)$,
and $A\cap L$ is connected for any line $L$ parallel to the $n$-th
axis, then 
\[
h_{\ominus\{0\}^{n-1}\times[0,1]}(A)=h(X^{n-1})+\log e.
\]
As a result, for general continuous random variables $X^{n}$ with
pdf $f$, 
\[
h_{\ominus\{0\}^{n}\times[0,1]}(\mathrm{hyp}_{+}f)=h(X^{n})+\log e,
\]
where $\mathrm{hyp}_{+}f=\left\{ (x,\alpha):\, x\in\mathbb{R}^{n},\,0<\alpha<f(x)\right\} \subseteq\mathbb{R}^{n+1}$. 
\end{enumerate}
\end{prop}
The proofs of these properties are given in Appendix~\ref{sub:erosion_prop}.

In the following proposition we show that $H(W_{A})$ can be upper bounded
using the erosion entropy. Moreover we show that the erosion entropy
is the average of the dyadic decomposition entropy under random shifting
and scaling. 
\begin{prop}
\label{prop:hdy_erosion}For a set $A\subseteq\mathbb{R}^{n}$ with a boundary of measure zero,
we have 
\[
H(W_{A})\le\log\mathrm{V}_{n}(A)+n h_{\ominus[0,1]^{n}}(A)+2n.
\]
Moreover, for any $T\in\mathbb{Z}$, $T>(1/n)\log\mathrm{V}_{n}(A)+1$,
when $U^{n}\sim\U[0,2^{T}]$ i.i.d., $\Theta\sim\U[0,1]$
independent of $U^{n}$ and $\Lambda=2^{\Theta}$, we have 
{
\normalfont
\[
\E\left[H(W_{\Lambda A+U})\right]=\log\mathrm{V}_{n}(A)+n h_{\ominus[0,1]^{n}}(A).
\]
}

\end{prop}
The proof of this proposition is given in Appendix~\ref{sub:hdy_erosion}.

For an orthogonally convex set $A$, the entropy of the dyadic decomposition
can be bounded by the volume of $A$ and the volume of its projection
as follows. 
\begin{thm}
\label{thm:ent_bound}Let $A\subseteq\mathbb{R}^{n}$ be an orthogonally convex set 
with $0<\mathrm{V}_{n}(A)<\infty$ and $X^{n}\sim\U(A)$,
then 
\[
H(W_{A})\le n\log\left(\sum_{i=1}^{n}\mathrm{VP}_{\backslash i}(A)\right)-\left(n-1\right)\log\mathrm{V}_{n}(A)+(2+\log e)n.
\]
Moreover, by applying the randomization in Proposition \ref{prop:hdy_erosion},
we obtain 
\begin{equation}
G(X_{1};\cdots;X_{n})\le n\log\left(\sum_{i=1}^{n}\mathrm{VP}_{\backslash i}(A)\right)-\left(n-1\right)\log\mathrm{V}_{n}(A)+n\log e.\label{eqn:gbound}
\end{equation}

\end{thm}
If the set $A$ is not orthogonally convex but can
be partitioned into orthogonally convex sets, then the property of
erosion entropy of union of sets in Proposition~\ref{prop:erosion_prop}
can be used to bound $H(W_{A})$. We now use the above theorem to
upper bound $G$ for the uniform pdf over an ellipse example.

\noindent \textbf{Example 1} (continued). Applying~\eqref{eqn:gbound}
to the uniform pdf over the ellipse $A=\left\{ x\in\mathbb{R}^{2}:\, x^{T}Kx\le1\right\} $,
we obtain 
\begin{align*}
H(W_{A}) & \le2\log\left(\sum_{i=1}^{2}\mathrm{VP}_{\backslash i}(A)\right)-\log\mathrm{V}_{2}(A)+4+2\log e\\
 & =2\log\left(2\sqrt{\frac{K_{11}}{\det K}}+2\sqrt{\frac{K_{22}}{\det K}}\right)-\log\left(\pi\sqrt{\frac{1}{\det K}}\right)+4+2\log e\\
 & =\log\left(\pi^{-1}\frac{\left(\sqrt{K_{11}}+\sqrt{K_{22}}\right)^{2}}{\sqrt{\det K}}\right)+6+2\log e.
\end{align*}
Comparing this to the mutual information for the uniform pdf over the
ellipse, we have 
\[
I(X_{1};X_{2})=\log\left(\pi e^{-1}\sqrt{\frac{K_{11}K_{22}}{\det K}}\right).
\]
Note that the gap between $H(W_{A})$ and $I(X_{1};X_{2})$ depends on the ratio
between $\left(\sqrt{K_{11}}+\sqrt{K_{22}}\right)^{2}$ and $\sqrt{K_{11}K_{22}}$,
which becomes very large when $K_{11}\gg K_{22}$. For example, if
$K=\diag(10000,1)$, then $\sqrt{K_{11}K_{22}}=100$ and
$I(X_{1};X_{2})\approx0.21$. On the other hand, $\left(\sqrt{K_{11}}+\sqrt{K_{22}}\right)^{2}=10201$
and the bound on $H(W_{A})$ is $13.02$. In the next section we show that this
gap can be reduced and bounded by a constant by appropriately 
scaling $A$. 

\medskip{}

To prove Theorem \ref{thm:ent_bound}, we need the following lemma,
which bounds the volume of the erosion of $A$ by a hypercube. 
\begin{lem}
\label{lem:ortho_onecon}For any orthogonally convex set $A\subseteq\mathbb{R}^{n}$
with $0<\mathrm{V}_{n}(A)<\infty$ and $\gamma\ge0$, the set $A\ominus\left([0,\gamma]\times\{0\}^{n-1}\right)\subseteq A$
is orthogonally convex, and 
\[
\mathrm{V}_{n}\left(A\ominus\left([0,\gamma]\times\{0\}^{n-1}\right)\right)\ge\mathrm{V}_{n}(A)-\int_{\mathrm{P}_{[2:n]}(A)}\min\left\{ \gamma\,,\,\mathrm{V}_{1}\left(A\cap(\mathrm{span}(\mathrm{e}_{1})+x)\right)\right\} dx_{2}^{n},
\]
where $\mathrm{span}(\mathrm{e}_{1})+x=\left\{ (\alpha,x_{2},x_{3},\ldots,x_{n})\,:\,\alpha\in\mathbb{R}\right\} \subseteq\mathbb{R}^{n}$.
As a result, 
\[
\mathrm{V}_{n}\left(A\ominus[0,\gamma]^{n}\right)\ge\mathrm{V}_{n}(A)-\sum_{i=1}^{n}\int_{\mathrm{P}_{\backslash i}(A)}\min\left\{ \gamma\,,\,\mathrm{V}_{1}\left(A\cap(\mathrm{span}(\mathrm{e}_{i})+x)\right)\right\} dx_{[1:n]\backslash i}.
\]
\end{lem}
\begin{IEEEproof}
We first prove the following result on the erosion of $A$ by a line
segment: for any orthogonally convex set $A\subseteq\mathbb{R}^{n}$
with $0<\mathrm{V}_{n}(A)<\infty$ and $\gamma\ge0$, the set $A\ominus\left([0,\gamma]\times\{0\}^{n-1}\right)\subseteq A$
is orthogonally convex, and {\allowdisplaybreaks 
\[
\mathrm{V}_{n}\left(A\ominus\left([0,\gamma]\times\{0\}^{n-1}\right)\right)\ge\mathrm{V}_{n}(A)-\int_{\mathrm{P}_{[2:n]}(A)}\min\left\{ \gamma\,,\,\mathrm{V}_{1}\left(A\cap(\mathrm{span}(\mathrm{e}_{1})+x)\right)\right\} \, dx.
\]
Note that 
\begin{align*}
A\ominus\left([0,\gamma]\times\{0\}^{n-1}\right) & =\left\{ x:\, x+\alpha\mathrm{e}_{1}\in A\;\text{for all}\;\alpha\in[0,\gamma]\right\} \\
 & =\left\{ x:\, x\in A-\alpha\mathrm{e}_{1}\;\text{for all}\;\alpha\in[0,\gamma]\right\} \\
 & =\bigcap_{\alpha\in[0,\gamma]}\left(A-\alpha\mathrm{e}_{1}\right)
\end{align*}
is the intersection of orthogonally convex sets, and therefore is
orthogonally convex. Also 
\begin{align*}
 & \mathrm{V}_{n}(A)-\mathrm{V}_{n}\left(A\ominus\left([0,\gamma]\times\{0\}^{n-1}\right)\right)\\
 & =\mathrm{V}_{n}\left\{ x\in A:\, x+\alpha\mathrm{e}_{1}\notin A\;\text{for some}\;\alpha\in[0,\gamma]\right\} \\
 & =\int_{\mathrm{P}_{[2:n]}(A)}\mathrm{V}_{1}\left\{ x_{1}\in\mathbb{R}:\,(x_{1},\tilde{x}_{2},\ldots,\tilde{x}_{n})\in A,\,(x_{1}+\alpha,\,\tilde{x}_{2},\ldots,\tilde{x}_{n})\notin A\;\text{for some}\;\alpha\in[0,\gamma]\right\} d\tilde{x}_{2}^{n}\\
 & =\int_{\mathrm{P}_{[2:n]}(A)}\mathrm{V}_{1}\left\{ x\in A\cap(\mathrm{span}(\mathrm{e}_{1})+\tilde{x}):\, x+\alpha\mathrm{e}_{1}\notin A\cap(\mathrm{span}(\mathrm{e}_{1})+\tilde{x})\;\text{for some}\;\alpha\in[0,\gamma]\right\} d\tilde{x}_{2}^{n}\\
 & \le\int_{\mathrm{P}_{[2:n]}(A)}\min\left\{ \gamma\,,\,\mathrm{V}_{1}\left(A\cap(\mathrm{span}(\mathrm{e}_{1})+\tilde{x})\right)\right\} d\tilde{x}_{2}^{n},
\end{align*}
} where the last inequality follows since $A\cap(\mathrm{span}(\mathrm{e}_{1})+\tilde{x})$
is connected.

By repeating this result for each axis, and observing that $\int_{\mathrm{P}_{\backslash i}(A)}\min\left\{ \gamma\,,\,\mathrm{V}_{1}\left(A\cap(\mathrm{span}(e_{i})+x)\right)\right\} dx_{[1:n]\backslash i}$
cannot increase when $A$ is replaced with an orthogonally convex
subset of $A$, we obtain the second bound. 
\end{IEEEproof}
We are now ready to prove Theorem \ref{thm:ent_bound}.

\medskip{}

\begin{IEEEproof}[Proof of Theorem \ref{thm:ent_bound}]
By Proposition \ref{prop:hdy_erosion}, the theorem can be proved
by bounding $h_{\ominus[0,1]^{n}}(A)$. Note that by Lemma \ref{lem:ortho_onecon},
\begin{align*}
  h_{\ominus[0,1]^{n}}(A)
 & =\int_{-\infty}^{\infty}\left(\mathbf{1}\left\{ t\ge0\right\} -\frac{\mathrm{V}_{n}\left(A\ominus[0,\,2^{-t}]^{n}\right)}{\mathrm{V}_{n}(A)}\right)dt\\
 & \le\int_{-\infty}^{\infty}\left(\mathbf{1}\left\{ t\ge0\right\} -\frac{1}{\mathrm{V}_{n}(A)}\max\left(0,\,\mathrm{V}_{n}(A)-\sum_{i=1}^{n}\int_{\mathrm{P}_{\backslash i}(A)}\min\left\{ 2^{-t}\,,\,\mathrm{V}_{1}\left(A\cap(\mathrm{span}(\mathrm{e}_{i})+x)\right)\right\} dx_{[1:n]\backslash i}\right)\right)dt\\
 & \le\int_{-\infty}^{\infty}\left(\mathbf{1}\left\{ t\ge0\right\} -\max\left(0,\,1-\frac{1}{\mathrm{V}_{n}(A)}\sum_{i=1}^{n}\int_{\mathrm{P}_{\backslash i}(A)}2^{-t}dx_{[1:n]\backslash i}\right)\right)dt\\
 & =\int_{-\infty}^{\infty}\left(\mathbf{1}\left\{ t\ge0\right\} -\max\left(0,\,1-2^{-t}\frac{\sum_{i=1}^{n}\mathrm{VP}_{\backslash i}(A)}{\mathrm{V}_{n}(A)}\right)\right)dt\\
 & =\log\left(\frac{\sum_{i=1}^{n}\mathrm{VP}_{\backslash i}(A)}{\mathrm{V}_{n}(A)}\right)+\log e.
\end{align*}
For the second result, note that the randomization in Proposition
\ref{prop:hdy_erosion} does not affect the right hand side of Theorem
\ref{thm:ent_bound}, which completes the proof of the theorem. 
\end{IEEEproof}
\medskip{}

\subsection{Scaling\label{subsec:scaling}}

In this section, we present a tighter bound on the common entropy
between continuous random variables by first scaling $A$ along each
dimension, that is, by performing a linear transformation $DA$ where $D$
is a diagonal matrix. This corresponds to scaling the random variable
$X_{i}$ by $D_{ii}$, $i\in[1:n]$, before applying the scheme. This new bound will
be in terms of the following. 
\begin{defn}[Truncated differential entropy]
Let $X^{n} \sim f(x^n)$ and define its truncated differential
entropy $\tilde{h}_{\zeta}(X^{n})$ for $\zeta\in(0,1]$, 
as 
\[
\tilde{h}_{\zeta}(X^{n})=\int_{\mathbb{R}^{n}}-\zeta^{-1}\min\left\{ \xi,\, f(x)\right\} \log\left(\zeta^{-1}\min\left\{ \xi,\, f(x)\right\} \right)dx,
\]
where $\xi>0$ such that 
\[
\int_{\mathbb{R}^{n}}\min\left\{ \xi,\, f(x)\right\} dx=\zeta.
\]
And define 
\[
\tilde{h}_{0}(X^{n})=\lim_{\zeta\to0}\tilde{h}_{\zeta}(X^{n})=\log\mathrm{V}_{n}\left\{ x:\, f(x)>0\right\} .
\]

\end{defn}
Note that $\tilde{h}_{\zeta}(X^{n})$ is decreasing in $\zeta$
from $\tilde{h}_{0}(X^{n})$ (the entropy of the uniform pdf
on the support of $X^{n}$) to $\tilde{h}_{1}(X^{n})=h(X^{n})$.

We now state the main result of this section, which shows that the
gap between $H(W_{DA})$ and $I_{\mathrm{D}}$ depends on how close
$\tilde{h}_{1/n}(X_{[1:n]\backslash i})$ and $h(X_{[1:n]\backslash i})$,
$i\in [1:n]$, are to each other. 
\begin{thm}
\label{thm:scale_min}For any orthogonally convex set $A\subseteq\mathbb{R}^{n}$
with $0<\mathrm{V}_{n}(A)<\infty$, there exists a diagonal matrix
$D\in\mathbb{R}^{n\times n}$ with positive diagonal entries such
that the entropy of the dyadic decomposition of $DA=\left\{ Dx:\, x\in A\right\} $
is bounded by 
\[
H(W_{DA})\le\sum_{i=1}^{n}\tilde{h}_{1/n}(X_{[1:n]\backslash i})-(n-1)\log\mathrm{V}_{n}(A)+n\log n+(2+\log e)n.
\]
Equivalently, when $X^{n}\sim\U(A)$, 
\[
H(W_{DA})\le I_{\mathrm{D}}(X_{1};\cdots;X_{n})+\sum_{i=1}^{n}\left(\tilde{h}_{1/n}(X_{[1:n]\backslash i})-h(X_{[1:n]\backslash i})\right)+n\log n+(2+\log e)n.
\]
Moreover, by applying the randomization in Proposition \ref{prop:hdy_erosion},
we obtain 
\[
I_{\mathrm{D}}\le J\le G\le I_{\mathrm{D}}+\sum_{i=1}^{n}\left(\tilde{h}_{1/n}(X_{[1:n]\backslash i})-h(X_{[1:n]\backslash i})\right)+n\log n+n\log e.
\]

\end{thm}
The proof of this theorem and a method for finding find $D$ are given in Appendix~\ref{sub:scale_min}.

We illustrate the above bound in the following.

\begin{figure}
\begin{centering}
\includegraphics[scale=0.42]{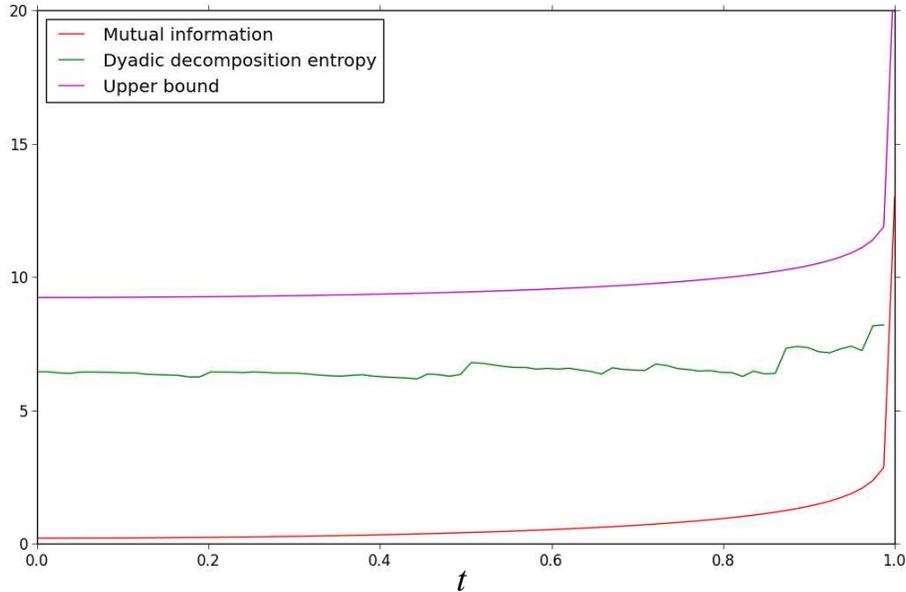} 
\par\end{centering}

\caption{\label{fig:ellipse_dyaent}Plot of the entropy of the dyadic decomposition,
mutual information, and the upper bound in Theorem \ref{thm:scale_min}
for Example \ref{exa:ellipse1}.}
\end{figure}

\noindent \textbf{Example 1} (continued). Applying Theorem \ref{thm:scale_min}
to the uniform pdf over the ellipse $A=\left\{ x\in\mathbb{R}^{2}:\, x^{T}Kx\le1\right\} $,
we have 
\begin{align*}
H(W_{DA}) & \le\sum_{i=1}^{2}\tilde{h}_{1/2}(X_{[1:2]\backslash i})-\log\mathrm{V}_{2}(A)+6+2\log e\\
 & \le\sum_{i=1}^{2}\log\left(\mathrm{VP}_{\backslash i}(A)\right)-\log\mathrm{V}_{2}(A)+6+2\log e\\
 & =\log\left(2\sqrt{\frac{K_{11}}{\det K}}\right)+\log\left(2\sqrt{\frac{K_{22}}{\det K}}\right)-\log\left(\pi\sqrt{\frac{1}{\det K}}\right)+6+2\log e\\
 & =\log\left(\pi^{-1}\sqrt{\frac{K_{11}K_{22}}{\det K}}\right)+8+2\log e.
\end{align*}
In comparison, the mutual information is
\[
I(X_{1};X_{2})=\log\left(\pi e^{-1}\sqrt{\frac{K_{11}K_{22}}{\det K}}\right),
\]
and the gap between $H(W_{DA})$ and $I(X_{1};X_{2})$
is bounded by a constant. Figure \ref{fig:ellipse_dyaent} plots the
values of $H(W_{DA})$ (calculated by finding all
squares in the dyadic decomposition with side length at least $2^{-11}$,
which yields a precise estimate), the upper bound in Theorem \ref{thm:scale_min},
and $I(X_{1};X_{2})$ for $K=\frac{1}{1-t^{2}}\left[\begin{array}{cc}
1 & -t\\
-t & 1
\end{array}\right]$, $t\in[0,1]$.

\section{Nonuniform Distributions\label{sec:logconcave}}

\begin{figure}[h]
\centering{}\includegraphics[scale=0.35]{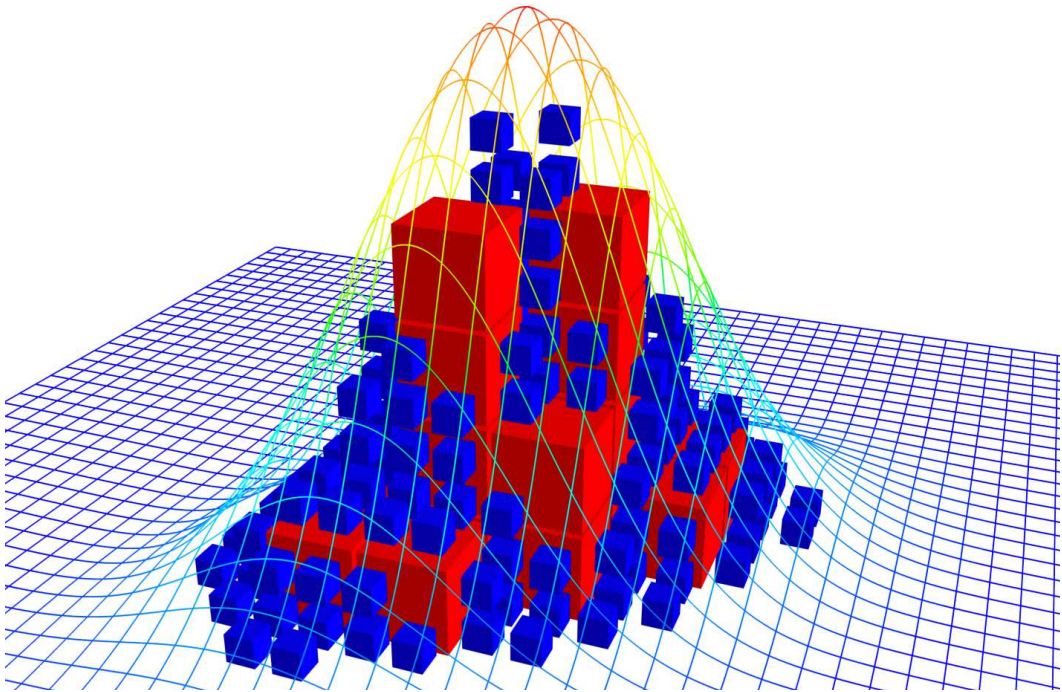} \caption{\label{fig:decomp_gauss}Dyadic decomposition of $\mathrm{hyp}_{+}f$
for the Gaussian pdf $f$ in Example~\ref{exa:gaussian}.}
\end{figure}

\begin{figure}[h]
\centering{}\includegraphics[scale=0.44]{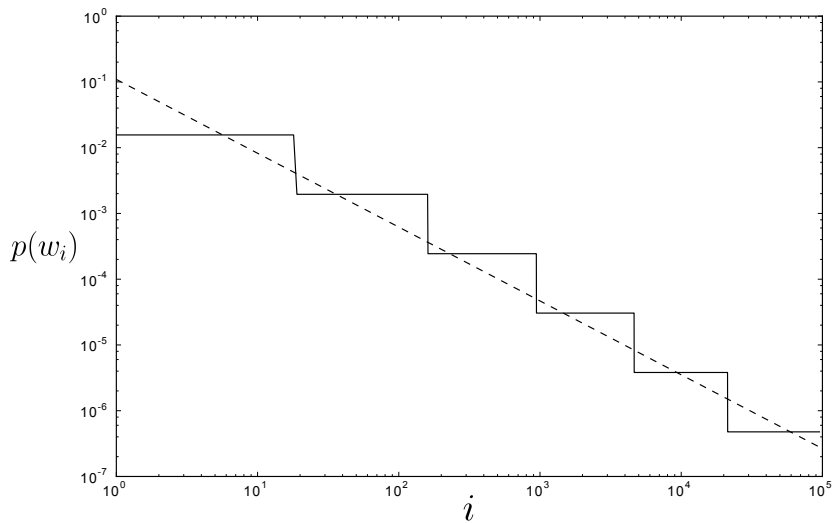} \caption{\label{fig:decomp_gauss_bar}The pmf of $W$ for the dyadic decomposition
of the hypograph of the Gaussian pdf in Example~\ref{exa:gaussian}.}
\end{figure}

In this section, we extend our results to the case in which the pdf of
$X^{n}$ is not necessarily uniform. Let $X^{n} \sim f(x^n)$
and let the support of $f$ be $A$. We add a random variable
$Z$ such that $(X_{1},\ldots,X_{n},Z)\sim\U(\mathrm{hyp}_{+}f)$,
where $\mathrm{hyp}_{+}f$ is the \emph{positive
strict hypograph} defined as 
\[
\mathrm{hyp}_{+}f=\left\{ (x,\alpha):\, x\in\mathbb{R}^{n},\,0<\alpha<f(x)\right\} \subseteq\mathbb{R}^{n+1}.
\]
Note that the marginal pdf of $X^n$ is $f$. Assuming that
$\mathrm{hyp}_{+}f$ is orthogonally convex, i.e., $f$ is orthogonally
concave, we can apply the results for the uniform pdf case in Section~\ref{sec:uniform}.
To illustrate this extension, consider the following. 
\begin{example}
\label{exa:gaussian} Let $(X_{1},X_{2})$ be zero mean Gaussian with
covariance matrix $K=\left[\begin{array}{cc}
1/8 & 1/16\\
1/16 & 1/8
\end{array}\right]$. Figure \ref{fig:decomp_gauss} plots the cubes with side length
$\ge 2^{-3}$ of the dyadic decomposition of the positive strict hypograph
of this pdf. Note that the cubes are scaled down so
as to show the ones behind them. Figure~\ref{fig:decomp_gauss_bar}
plots the pmf of $W_{\mathrm{hyp}_{+}f}$ in log-log scale. As in
Example~\ref{exa:ellipse1}, the tail of the pmf  follows a power law with power
approximately $1.12$ and $H(W_{\mathrm{hyp}_{+}f})$ is finite. 
\end{example}

We now show that if $f$ is log-concave (i.e., $x\mapsto\log f(x)$
is concave), then the difference between the entropy of the dyadic
decomposition and the dual total correlation is bounded by a constant
that depends only on $n$. 
\begin{thm}
\label{thm:logconcave}If the pdf of $X^{n}$ is log-concave, then
there exists a diagonal matrix $D\in\mathbb{R}^{(n+1)\times(n+1)}$
with positive diagonal entries such that the entropy of the dyadic
decomposition of $D\mathrm{hyp}_{+}f$ satisfies 
\[
\begin{aligned}H(W_{D\mathrm{hyp}_{+}f}) & \le I_{\mathrm{D}}(X_{1};\cdots;X_{n})+n^{2}\log e+n\left(\log n+\log(n+1)+e+2\log e+2\right)+2+\log e\\
 & \le I_{\mathrm{D}}(X_{1};\cdots;X_{n})+n^{2}\log e+12n\log n.
\end{aligned}
\]
Moreover, by applying the randomization in Proposition \ref{prop:hdy_erosion},
we obtain 
\[
I_{\mathrm{D}}\le J\le G\le I_{\mathrm{D}}+n^{2}\log e+9n\log n.
\]

\end{thm}
To prove Theorem \ref{thm:logconcave}, we first state a result in
 \cite{bobkov2011entropy}, which bounds the differential entropy
of a log-concave pdf by its maximum density. 
\begin{lem}
\label{lem:logconcave_sup}For any log-concave pdf $f$ of $X^{n}$,
\[
-\log\Bigl(\sup_{x\in\mathbb{R}^{n}}f(x)\Bigr)\le h(X^{n})\le-\log\Bigl(\sup_{x\in\mathbb{R}^{n}}f(x)\Bigr)+n\log e.
\]

\end{lem}
We can generalize Lemma \ref{lem:logconcave_sup} to bound the conditional
differential entropy as follows. 
\begin{lem}
\label{lem:logconcave_cond}For any log-concave pdf $f$ of $X^{n}$,
and $1\le m\le n$, 
\[
0 \le h(X_{m+1}^{n}\,|\, X^{m})+\log\Bigl(\int_{\mathbb{R}^{m}}\sup_{\tilde{x}_{m+1}^{n}\in\mathbb{R}^{n-m}}f(x^{m},\,\tilde{x}_{m+1}^{n})dx^{m}\Bigr) \le n\log e+\log\binom{n}{m}.
\]

\end{lem}
The proof of this lemma is given in Appendix~\ref{sub:logconcave_cond}.
\medskip{}

Next, we establish a bound on the difference between the differential
entropy and the truncated differential entropy. \medskip{}

\begin{lem}
\label{lem:logconcave_trunc}For any log-concave pdf $f$, 
\[
\tilde{h}_{\zeta}(X^{n})-h(X^{n})\le\log\zeta+\nu+n\log e,
\]
where $\nu\ge0$ satisfies 
\[
\Gamma(n+1,\,\nu)=\zeta\cdot\Gamma(n+1),
\]
and $\Gamma(n,z)=\int_{z}^{\infty}t^{n-1}e^{-t}dt$ is the incomplete
gamma function, and $\Gamma(n)=\Gamma(n,0)$ is the gamma function.
Moreover, if $\zeta\ge e^{-(e-2)n}$, then 
\[
\tilde{h}_{\zeta}(X^{n})-h(X^{n})\le\log\zeta+(e+\log e)n.
\]

\end{lem}
The proof of this lemma is given in Appendix~\ref{sub:logconcave_trunc}.

\medskip{}

We now proceed to the proof of Theorem \ref{thm:logconcave}.

\medskip{}

\begin{IEEEproof}[Proof of Theorem \ref{thm:logconcave}]
Let $(X^n,Z)\sim\U(\mathrm{hyp}_{+}f)$.
Applying Theorem \ref{thm:scale_min} on $\mathrm{hyp}_{+}f$, we
have 
\[
\begin{aligned}H(W_{D\mathrm{hyp}_{+}f}) & \le\tilde{h}_{1/(n+1)}(X^{n})+\sum_{i=1}^{n}\tilde{h}_{1/(n+1)}(X_{[1:n]\backslash i},\, Z)+(n+1)\log(n+1)+(2+\log e)(n+1)\\
 & \le\tilde{h}_{1/(n+1)}(X^{n})+\sum_{i=1}^{n}\log\left(\mathrm{VP}_{\backslash i}(\mathrm{hyp}_{+}f)\right)+(n+1)\log(n+1)+(2+\log e)(n+1).
\end{aligned}
\]
Consider the term $\tilde{h}_{1/(n+1)}(X^{n})$. Since $1/(n+1)\ge e^{-(e-2)n}$, by Lemma (\ref{lem:logconcave_trunc}) we have 
\[
\tilde{h}_{1/(n+1)}(X^{n})-h(X^{n})\le-\log(n+1)+(e+\log e)n.
\]
Consider the term $\log\left(\mathrm{VP}_{\backslash n}(\mathrm{hyp}_{+}f)\right)$.
By Lemma (\ref{lem:logconcave_cond}), we have 
\[
\begin{aligned} & \log\left(\mathrm{VP}_{\backslash n}(\mathrm{hyp}_{+}f)\right)\\
 & =\log\int_{\mathbb{R}^{n-1}}\sup_{\tilde{x}_{n}}f(x_{1}^{n-1},\tilde{x}_{n})dx_{1}^{n-1}\\
 & \le-h(X_{n}|X_{1}^{n-1})+n\log e+\log\binom{n}{n-1}\\
 & =  -h(X_{n}|X_{1}^{n-1})+n\log e+\log n.
\end{aligned}
\]
Hence 
\[
\begin{aligned} & H(W_{D\mathrm{hyp}_{+}f})\\
 & \le h(X^{n})-\log(n+1)+(e+\log e)n\\
 & \;\;+\sum_{i=1}^{n}\left(-h(X_{i}|X_{[1:n]\backslash i})+n\log e+\log n\right)+(n+1)\log(n+1)+(2+\log e)(n+1)\\
 & =I_{\mathrm{D}}(X_{1};\cdots;X_{n})+n^{2}\log e+n\left(\log n+\log(n+1)+e+2\log e+2\right)+2+\log e.
\end{aligned}
\]

\end{IEEEproof}

Note that the result in Theorem~\ref{thm:logconcave} can be readily extended to mixtures of log-concave pdfs using the union property in Proposition~\ref{prop:erosion_prop}. It is not possible to obtain a constant bound on the gap between $I_{\mathrm{D}}$ and $G$ for arbitrary pdfs, however. To\\[3pt] see this, 
let $(\tilde{X}_1,\tilde{X}_2) \in \{1,\ldots,2^m\}^2$ be two discrete random variables with pmf
$\begin{bmatrix}
1/3 & 1/3 \\ 
1/3 & 0
\end{bmatrix}^{\otimes m}$,
i.e., $(\tilde{X}_1,\tilde{X}_2)$ consists of $m$ i.i.d. copies of two random variables with pmf
$\begin{bmatrix}
1/3 & 1/3 \\ 
1/3 & 0
\end{bmatrix}$.
Now let $X_1=\tilde{X}_1+Z_1$ and  $X_2=\tilde{X}_2+Z_2$, where\\[3pt] $Z_1,Z_2 \sim \U [0,1]$. Then we have $I(X_1;X_2) = I(\tilde{X}_1;\tilde{X}_2) = (\log 3 - 4/3)m$, $J(X_1;X_2) = J(\tilde{X}_1;\tilde{X}_2) = (2/3)m$, and the gap between $I$ and $J$ grows linearly in $m$. Since $G\ge J$, the gap between $I$ and $G$ grows at least linearly in $m$.

\section{Approximate Distributed Simulation\label{sec:approx_bdd}}

The cardinality of $W$ for exact simulation of $n$
continuous random variables is in general infinite, hence the length of the codeword for 
$W$ is unbounded. We show that if the exact simulation requirement
is relaxed by only requiring that the total variation
distance between the distributions of the simulated and the prescribed
random variables to be small, then distributed simulation is possible with a fixed
length code.

We define the approximate distributed simulation problem as follows.
There are $n$ agents that have access to common randomness $W\in\{0,1\}^{N}$. Agent $i\in[1:n]$ wishes to simulate the random
variable $\tilde{X}_{i}$ using $W$ and its local randomness, which
is independent of $W$ and local randomness at other agents, such
that the total variation between the distributions of $\tilde{X}^n$ 
and $X^n$ is bounded as
\[
d_{\mathrm{TV}}\left((\tilde{X}_{1},\ldots,\tilde{X}_{n})\,,\,(X_{1},\ldots,X_{n})\right)\le\epsilon,
\]
for some $\epsilon>0$. The problem is to find the conditions under which
the length-distance pair $(N,\epsilon)$ is achievable.

We can find sufficient conditions under which $(N,\epsilon)$ is achievable by terminating the dyadic decomposition
scheme described in the previous sections after a finite number of iterations, that is, by discarding all hypercubes smaller than
a prescribed size. The following proposition gives the length-distance
pairs achievable by this truncated dyadic decomposition scheme
in terms of $H(W_A)$ for uniform pdf over $A$ (or $H(W_{D\mathrm{hyp}_{+}f})$ for non-uniform pdfs), which in turn can be bounded by Theorem~\ref{thm:ent_bound}, \ref{thm:scale_min} or \ref{thm:logconcave}.
\begin{thm}
\label{thm:approx_logconcave}Let $X^n \sim \U(A)$,
where $A\subseteq\mathbb{R}^{n}$ with a boundary of measure zero. The truncated dyadic decomposition
scheme can achieve the length-distance pair $(N,\epsilon)$ if 
\[
N\ge\log\left(\epsilon2^{\epsilon^{-1}H(W_{A})}+1\right).
\]
\end{thm}
\begin{IEEEproof}
Let $W_{A}$ be the dyadic decomposition common randomness variable for $X^n\U(A)$ and define 
\[
\tilde{W}_{A}=\begin{cases}
(k,v) & \text{if}\; k<l\\
(k_{0},v_{0}) & \text{if}\; k\ge l,
\end{cases}
\]
where $W_{A}=(k,v)$, and $(k_{0},v_{0})$ is any hypercube with side length $> 2^{-l}$, where $l=n^{-1}(\epsilon^{-1}H(W_{A})-\log\mathrm{V}_{n}(A))$. The truncated scheme uses $\tilde{W}_{A}$ as common
randomness variable, and the operations performed by the
agents are unchanged. Since the truncated scheme differs from the
original scheme only when $\tilde{W}_{A}\neq W_{A}$, we have
\[
\begin{aligned} & d_{\mathrm{TV}}\left((\tilde{X}_{1},\ldots,\tilde{X}_{n})\,,\,(X_{1},\ldots,X_{n})\right)\\
 & \le\P\left\{ K\ge l\right\} \\
 & =\P\left\{ \log\mathrm{V}_{n}(A)+nK\ge\log\mathrm{V}_{n}(A)+nl\right\} \\
 & \le\frac{\E\left[\log\mathrm{V}_{n}(A)+nK\right]}{\log\mathrm{V}_{n}(A)+nl}\\
 & =\frac{H(W_{A})}{\log\mathrm{V}_{n}(A)+nl}\\
 & =\epsilon.
\end{aligned}
\]
It is left to bound the cardinality of $\tilde{W}_{A}$. Consider the
probability vector $p_{\tilde{W}_{A}}\in\mathbb{R}^{|\tilde{W}_{A}|}$.
Since $p_{\tilde{W}_{A}}(w)\ge2^{-nl}\mathrm{V}_{n}^{-1}(A)$, the
vector $p_{\tilde{W}_{A}}$ can be expressed as a convex combination
of the vectors $2^{-nl}\mathrm{V}_{n}^{-1}(A)\mathbf{1}+\left(1-2^{-nl}\mathrm{V}_{n}^{-1}(A)|\tilde{W}_{A}|\right)\mathrm{e}_{i}$
for $i=1,\ldots,|\tilde{W}_{A}|$. We have 
\[
\begin{aligned} & H\left(2^{-nl}\mathrm{V}_{n}^{-1}(A)\mathbf{1}+\left(1-2^{-nl}\mathrm{V}_{n}^{-1}(A)|\tilde{W}_{A}|\right)\mathrm{e}_{i}\right)\\
 & =2^{-nl}\mathrm{V}_{n}^{-1}(A)(|\tilde{W}_{A}|-1)\cdot\left(\log\mathrm{V}_{n}(A)+nl\right)-\left(1-2^{-nl}\mathrm{V}_{n}^{-1}(A)(|\tilde{W}_{A}|-1)\right)\log\left(1-2^{-nl}\mathrm{V}_{n}^{-1}(A)(|\tilde{W}_{A}|-1)\right)\\
 & \ge2^{-nl}\mathrm{V}_{n}^{-1}(A)(|\tilde{W}_{A}|-1)\cdot\left(\log\mathrm{V}_{n}(A)+nl\right).
\end{aligned}
\]
Since entropy is concave, $H(W_{A})\ge H(\tilde{W}_{A})\ge2^{-nl}\mathrm{V}_{n}^{-1}(A)(|\tilde{W}_{A}|-1)\cdot\left(\log\mathrm{V}_{n}(A)+nl\right)$.
By Theorem~\ref{thm:logconcave},
\[
\begin{aligned}|\tilde{W}_{A}| & \le\frac{1}{\log\mathrm{V}_{n}(A)+nl}2^{nl}\mathrm{V}_{n}(A)H(W_{A})+1\\
 & =\frac{1}{\epsilon^{-1}H(W_{A})}2^{\epsilon^{-1}H(W_{A})}H(W_{A})+1\\
 & =\epsilon2^{\epsilon^{-1}H(W_{A})}+1\\
 & \le2^{N}.
\end{aligned}
\]
The result follows. 
\end{IEEEproof}

\section{Conclusion\label{sec:conclusion}}

We proposed a scheme for distributed simulation of continuous random variables
based on dyadic decomposition. We established a bound on the entropy of the constructed common randomness
in terms of the dual total correlation for the class of log concave pdfs. As a result, the gap between exact and
Wyner's common information and dual total correlation can be bounded for this set of distributions.

Our results readily translate to the exact, one-shot
version of the channel synthesis problem in~\cite{cuff2010coordination,cuff2013distributed}
without common randomness in which we wish to simulate a channel $f_{Y|X}(y|x)$
with input distribution $f_{X}(x)$. Given the input $X\sim f_{X}$,
the encoder produces the codeword $W$ using a prefix-free code. Upon
receiving $W$, the decoder produces the output $\hat{Y}$ such that
$(X,\hat{Y})\sim f_{X}f_{Y|X}$. The problem again is to find the 
minimum entropy of $W$. A consequence of our results is that an additive
Gaussian noise channel with Gaussian input, can be exactly simulated using only a finite amount of common randomness.

We have seen in Section~\ref{subsec:scaling} that
performing different scalings on each $X_i$ can reduce $H(W)$. More generally, applying a bijective transformation  $g_i(x_i)$ to each random variable before using the dyadic decomposition scheme may help reduce $H(W)$ further. For example, applying the copula transform~\cite{sklar1959fonctions} $g_i(x)=F_{X_i}(x)$ such that $g_i(X_i)\sim \U[0,1]$ has the benefit that when the $X_i$'s are close to independent, the pdf is close to a constant function over the unit hypercube, which is likely to result in a smaller $H(W)$. 

Finally, our results readily apply to the distributed randomness generation setting in which the agents share a stream of uniformly random bits instead of a codeword generated by an active encoder. In this setting, the agents would need to agree on the number of random bits used (i.e., they recover $W$ from the stream using the an optimal prefix-free code). In this case, the optimal expected number of random bits used is between $H(\tilde{W})$ and $H(\tilde{W})+2$ (see~\cite{knuth1976complexity}). Hence it is also sufficient to consider $H(W)$.

\appendix
{}

\subsection{Bounding Common Information by Dual Total Correlation\label{sub:jd_ci}}

We show that $I_{\mathrm{D}}(X_{1};\cdots;X_{n})\le J(X_{1};\cdots;X_{n})$.
For general random variables, the dual total correlation
is defined as 
\[
I_{\mathrm{D}}(X_{1};\cdots;X_{n})=\sum_{i=1}^{n-1}I(X_{i};X_{i+1}^{n}\,|\, X_{1}^{i-1}).
\]
To prove the inequality, let $W$ be a random variable such that $X^{n}$
are conditionally independent given $W$, then
\[
\begin{aligned}I(W;X^{n}) & =I(X_{1};X_{2}^{n})+I(W;X_{2}^{n}|X_{1})+I(W;X_{1}|X_{2}^{n})\\
 & \ge I(X_{1};X_{2}^{n})+I(W;X_{2}^{n}|X_{1})\\
 & =I(X_{1};X_{2}^{n})+I(X_{2};X_{3}^{n}|X_{1})+I(W;X_{3}^{n}|X_{1}^{2})+I(W;X_{2}|X_{1},X_{3}^{n})\\
 & \ge I(X_{1};X_{2}^{n})+I(X_{2};X_{3}^{n}|X_{1})+I(W;X_{3}^{n}|X_{1}^{2})\\
 & \vdots\\
 & \ge I(X_{1};X_{2}^{n})+I(X_{2};X_{3}^{n}|X_{1})+\cdots+I(X_{n-1};X_{n}|X_{1}^{n-2}).
\end{aligned}
\]

\subsection{Dyadic Decomposition Algorithm Details\label{sec:schemes}}

We present the common randomness generation and simulation algorithms
for $X^{n}\sim\U(A)$ using arithmetic coding. The common random
source runs the generation algorithm to produce $W$ and agent $i\in[1:n]$
runs the simulation algorithm with input $W$ to produce $x_{i}$.

\medskip{}
 \textbf{Common randomness generation algorithm.}

\noindent Input: $A\subseteq[0,1]^{n}$\\
Output: codeword $w$ 
\begin{enumerate}
\item $v\leftarrow(0,\ldots,0)$, $k\leftarrow0$, $\alpha\leftarrow0$,
$(\mu,\nu)\leftarrow(0,1)$, $w\leftarrow\emptyset\in\{0,1\}^{*}$
\item While $C_{k,v}=2^{-k}([0,1]^{n}+v)\nsubseteq A$: 
\item $\quad\quad$For each $\tilde{v}\in\{0,1\}^{n}+2v=\{2v_{1},2v_{1}+1\}\times\cdots\times\{2v_{n},2v_{n}+1\}$: 
\item $\quad\quad\quad\quad$$p_{\tilde{v}}\leftarrow\mathrm{V}_{n}(A\cap C_{k+1,\tilde{v}})/\mathrm{V}_{n}(A)$
\item $\quad\quad$While $[\mu,\nu]\nsubseteq[\alpha+\sum_{\hat{v}\prec\tilde{v}}p_{\hat{v}},\,\alpha+\sum_{\hat{v}\preceq\tilde{v}}p_{\hat{v}}]$
for all $\tilde{v}\in\{0,1\}^{n}+2v$ ($\prec$ is lexicographical
order)
\item $\quad\quad\quad\quad$Generate $\tilde{w}\in\{0,1\}$ uniformly at
random
\item $\quad\quad\quad\quad$$w\leftarrow w\,\Vert\,\tilde{w}$ (append
$\tilde{w}$ to $w$) 
\item $\quad\quad\quad\quad$$(\mu,\nu)\leftarrow(\mu+(\nu-\mu)\tilde{w}/2,\,\mu+(\nu-\mu)(\tilde{w}+1)/2)$
\item $\quad\quad$$v\leftarrow\tilde{v}$ where $[\mu,\nu]\subseteq[\alpha+\sum_{\hat{v}\prec\tilde{v}}p_{\hat{v}},\,\alpha+\sum_{\hat{v}\preceq\tilde{v}}p_{\hat{v}}]$
\item $\quad\quad$$\alpha\leftarrow\alpha+\sum_{\hat{v}\prec\tilde{v}}p_{\hat{v}}$
\item Output $w$

\medskip{}

\end{enumerate}
\noindent \textbf{Simulation algorithm.}

Input: Agent $i$, common randomness $w$,
$A\subseteq[0,1]^{n}$

Output: random variate $x_{i}$ 
\begin{enumerate}
\item $v\leftarrow(0,\ldots,0)$, $k\leftarrow0$, $\alpha\leftarrow0$,
$(\mu,\nu)\leftarrow(0,1)$
\item While $C_{k,v}=2^{-k}([0,1]^{n}+v)\nsubseteq A$: 
\item $\quad\quad$For each $\tilde{v}\in\{0,1\}^{n}+2v=\{2v_{1},2v_{1}+1\}\times\cdots\times\{2v_{n},2v_{n}+1\}$: 
\item $\quad\quad\quad\quad$$p_{\tilde{v}}\leftarrow\mathrm{V}_{n}(A\cap C_{k+1,\tilde{v}})/\mathrm{V}_{n}(A)$
\item $\quad\quad$While $[\mu,\nu]\nsubseteq[\alpha+\sum_{\hat{v}\prec\tilde{v}}p_{\hat{v}},\,\alpha+\sum_{\hat{v}\preceq\tilde{v}}p_{\hat{v}}]$
for all $\tilde{v}\in\{0,1\}^{n}+2v$ ($\prec$ is lexicographical
order)
\item $\quad\quad\quad\quad$$\tilde{w}\leftarrow$first bit of $w$, discard
first bit of $w$
\item $\quad\quad\quad\quad$$(\mu,\nu)\leftarrow(\mu+(\nu-\mu)\tilde{w}/2,\,\mu+(\nu-\mu)(\tilde{w}+1)/2)$
\item $\quad\quad$$v\leftarrow\tilde{v}$ where $[\mu,\nu]\subseteq[\alpha+\sum_{\hat{v}\prec\tilde{v}}p_{\hat{v}},\,\alpha+\sum_{\hat{v}\preceq\tilde{v}}p_{\hat{v}}]$
\item $\quad\quad$$\alpha\leftarrow\alpha+\sum_{\hat{v}\prec\tilde{v}}p_{\hat{v}}$
\item Output randomly generated $x_{i}$ according to $\U[2^{-k}v_{i},\,2^{-k}(v_{i}+1)]$ 
 \medskip{}

\end{enumerate}
The above algorithms assume $A\subseteq[0,1]^{n}$. The case where
$A$ is unbounded can be handled by first encoding the integer parts
$\left\lfloor X^{n}\right\rfloor =(\left\lfloor X_{1}\right\rfloor ,\ldots,\left\lfloor X_{n}\right\rfloor )$
of $X^{n}\sim\U(A)$, then run the algorithms on $A\cap([0,1]^{n}+\left\lfloor X^{n}\right\rfloor )$.
The algorithms can be applied to non-uniform pdfs by letting $A$
to be $\mathrm{hyp}_{+}f$ scaled according to Theorem \ref{thm:scale_min}.

An advantage of the generation and simulation algorithms based on
arithmetic coding is that each bit of $w$ is generated uniformly,
and hence can be applied to the situation where the agents share a
stream of uniformly random bits \cite{knuth1976complexity}.

\subsection{Proof of Proposition \ref{prop:erosion_prop} \label{sub:erosion_prop}}

The monotonicity property and the linear transformation property follow
directly from the definition of erosion entropy.

For the scaling property, consider
\begin{align*}
h_{\ominus\beta B}(\alpha A) & =\int_{-\infty}^{\infty}\left(\mathbf{1}\left\{ t\ge0\right\} -\frac{\mathrm{V}_{n}\left(\alpha A\ominus2^{-t}\beta B\right)}{\mathrm{V}_{n}(\alpha A)}\right)dt\\
 & =\int_{-\infty}^{\infty}\left(\mathbf{1}\left\{ t\ge0\right\} -\frac{\mathrm{V}_{n}\left(A\ominus2^{-t+\log(\beta/\alpha)}B\right)}{\mathrm{V}_{n}(A)}\right)dt\\
 & =\int_{-\infty}^{\infty}\left(\mathbf{1}\left\{ t\ge-\log(\beta/\alpha)\right\} -\frac{\mathrm{V}_{n}\left(A\ominus2^{-t}B\right)}{\mathrm{V}_{n}(A)}\right)dt\\
 & \ge h_{\ominus B}(A)+\log(\beta/\alpha).
\end{align*}

For the union property, consider
\begin{align*}
h_{\ominus B}\left(\bigcup_{i=1}^{k}A_{i}\right) & =\int_{-\infty}^{\infty}\left(\mathbf{1}\left\{ t\ge0\right\} -\frac{\mathrm{V}_{n}\left(\left(\bigcup_{i}A_{i}\right)\ominus2^{-t}B\right)}{\sum_{i}\mathrm{V}_{n}(A_{i})}\right)dt\\
 & \le\int_{-\infty}^{\infty}\left(\mathbf{1}\left\{ t\ge0\right\} -\frac{\mathrm{V}_{n}\left(\bigcup_{i}\left(A_{i}\ominus2^{-t}B\right)\right)}{\sum_{i}\mathrm{V}_{n}(A_{i})}\right)dt\\
 & =\int_{-\infty}^{\infty}\left(\mathbf{1}\left\{ t\ge0\right\} -\frac{\sum_{i}\mathrm{V}_{n}\left(A_{i}\ominus2^{-t}B\right)}{\sum_{i}\mathrm{V}_{n}(A_{i})}\right)dt\\
 & =\int_{-\infty}^{\infty}\left(\sum_{i}\frac{\mathrm{V}_{n}(A_{i})}{\sum_{j}\mathrm{V}_{n}(A_{j})}\left(\mathbf{1}\left\{ t\ge0\right\} -\frac{\mathrm{V}_{n}\left(A_{i}\ominus2^{-t}B\right)}{\mathrm{V}_{n}(A_{i})}\right)\right)dt\\
 & =\sum_{i}\frac{\mathrm{V}_{n}(A_{i})}{\sum_{j}\mathrm{V}_{n}(A_{j})}\cdot h_{\ominus B}(A_{i}).
\end{align*}
Equality holds if $\left(\bigcup_{i}A_{i}\right)\ominus2^{-t}B=\bigcup_{i}\left(A_{i}\ominus2^{-t}B\right)$,
which is true when the closures of $A_{1},\ldots,A_{k}$ are disjoint.

For the reduction to differential entropy property, let $X^{n}\sim\U(A)$,
and $A\cap L$ is connected for any line $L$ parallel to the $n$-th
axis, then
\begin{align*}
h_{\ominus\{0\}^{n-1}\times[0,1]}(A) & =\int_{-\infty}^{\infty}\left(\mathbf{1}\left\{ t\ge0\right\} -\frac{\mathrm{V}_{n}\left(A\ominus\left(\{0\}^{n-1}\times[0,2^{-t}]\right)\right)}{\mathrm{V}_{n}(A)}\right)dt\\
 & =\int_{-\infty}^{\infty}\left(\mathbf{1}\left\{ t\ge0\right\} -\int_{\mathbb{R}^{n-1}}\frac{\mathrm{V}_{1}\left(\left(A\ominus\{0\}^{n-1}\times[0,2^{-t}]\right)\cap\left(\{x_{1}^{n-1}\}\times\mathbb{R}\right)\right)}{\mathrm{V}_{n}(A)}dx_{1}^{n-1}\right)dt\\
 & =\int_{-\infty}^{\infty}\left(\mathbf{1}\left\{ t\ge0\right\} -\int_{\mathbb{R}^{n-1}}\max\left\{ f_{X_{1}^{n-1}}(x_{1}^{n-1})-2^{-t},\,0\right\} dx_{1}^{n-1}\right)dt\\
 & =\int_{\mathbb{R}^{n-1}}f_{X_{1}^{n-1}}(x_{1}^{n-1})\int_{-\infty}^{\infty}\left(\mathbf{1}\left\{ t\ge0\right\} -\max\left\{ 1-2^{-t}/f_{X_{1}^{n-1}}(x_{1}^{n-1}),\,0\right\} \right)dtdx_{1}^{n-1}\\
 & =\int_{\mathbb{R}^{n-1}}f_{X_{1}^{n-1}}(x_{1}^{n-1})\int_{-\infty}^{\infty}\left(\mathbf{1}\left\{ t\ge0\right\} -\max\left\{ 1-2^{-t}/f_{X_{1}^{n-1}}(x_{1}^{n-1}),\,0\right\} \right)dtdx_{1}^{n-1}\\
 & =\int_{\mathbb{R}^{n-1}}f_{X_{1}^{n-1}}(x_{1}^{n-1})\left(-\log f_{X_{1}^{n-1}}(x_{1}^{n-1})+\int_{-\infty}^{\infty}\left(\mathbf{1}\left\{ t\ge0\right\} -\max\left\{ 1-2^{-t},\,0\right\} \right)dtdx_{1}^{n-1}\right)\\
 & =h(X_{1}^{n-1})+\log e.
\end{align*}

\subsection{Proof of Proposition \ref{prop:hdy_erosion} \label{sub:hdy_erosion}}

Note that{\allowdisplaybreaks
\begin{align*}
\sum_{l=-\infty}^{k}2^{-nl}\left|D_{l}(A)\right|  =2^{-nk}\left|\left\{ v\in\mathbb{Z}^{n}:\, C_{k,v}\subseteq A\right\} \right|
  \ge2^{-nk}\left|\left\{ v\in\mathbb{Z}^{n}:\, C_{k-1,(v-w)/2}\subseteq A\right\} \right|.
\end{align*}
for any $w\in[0,1]^{n}$, since $C_{k,v}\subseteq C_{k-1,(v-w)/2}$.}
Note that the $(v-w)/2$ in the subscript may not have integer entries,
but still the same definition $C_{k,v}=2^{-k}([0,1]^{n}+v)$ can be
applied. Also{\allowdisplaybreaks 
\begin{align*}
  \int_{[0,1]^{n}}\left|\left\{ v\in\mathbb{Z}^{n}:\, C_{k-1,(v-w)/2}\subseteq A\right\} \right| dw
 & =\sum_{v\in\mathbb{Z}^{n}}\int_{[0,1]^{n}}\mathbf{1}\left\{ C_{k-1,(v-w)/2}\subseteq A\right\} dw\\
 & =2^{n}\int_{\mathbb{R}^{n}}\mathbf{1}\left\{ C_{k-1,w}\subseteq A\right\} dw\\
 & =2^{n}2^{n(k-1)}\mathrm{V}_{n}\left(A\ominus[0,\,2^{-(k-1)}]^{n}\right)\\
 & =2^{nk}\mathrm{V}_{n}\left(A\ominus[0,\,2^{-(k-1)}]^{n}\right).
\end{align*}
Hence 
\[
\sum_{l=-\infty}^{k}2^{-nl}\left|D_{l}(A)\right|\ge\mathrm{V}_{n}\left(A\ominus[0,\,2^{-(k-1)}]^{n}\right),
\]
\[
\sum_{l=k+1}^{\infty}2^{-nl}\left|D_{l}(A)\right|\le\mathrm{V}_{n}(A)-\mathrm{V}_{n}\left(A\ominus[0,\,2^{-(k-1)}]^{n}\right).
\]
} Note that $H(W_{A})=H(W_{(1/2)A})$, and also the right-hand-side
of the proposition remains the same when $A$ is replaced
by $(1/2)A$. Without loss of generality assume $A$ is small enough
such that $\mathrm{V}_{n}(A)\le1$, so $D_{k}(A)=\emptyset$ for $k<0$. 
\begin{align*}
H(W_{A}) & =\log\mathrm{V}_{n}(A)+\frac{1}{\mathrm{V}_{n}(A)}\sum_{k=0}^{\infty}nk2^{-nk}\left|D_{k}(A)\right|\\
 & =\log\mathrm{V}_{n}(A)+\frac{n}{\mathrm{V}_{n}(A)}\sum_{k=0}^{\infty}\sum_{l=k+1}^{\infty}2^{-nl}\left|D_{l}(A)\right|\\
 & \le\log\mathrm{V}_{n}(A)+\frac{n}{\mathrm{V}_{n}(A)}\sum_{k=0}^{\infty}\left(\mathrm{V}_{n}(A)-\mathrm{V}_{n}\left(A\ominus[0,\,2^{-(k-1)}]^{n}\right)\right)\\
 & \le\log\mathrm{V}_{n}(A)+\frac{n}{\mathrm{V}_{n}(A)}\int_{-2}^{\infty}\left(\mathrm{V}_{n}(A)-\mathrm{V}_{n}\left(A\ominus[0,\,2^{-t}]^{n}\right)\right)dt\\
 & =\log\mathrm{V}_{n}(A)+n\cdot h_{\ominus[0,1]^{n}}(A)+2n.
\end{align*}

To prove the second result, consider 
\begin{align*}
\sum_{l=-\infty}^{k}2^{-nl}\left|D_{l}\left(\Lambda(A+U)\right)\right| & =2^{-nk}\left|\left\{ v\in\mathbb{Z}^{n}:\, C_{k,v}\subseteq\Lambda(A+U)\right\} \right|\\
 & =2^{-nk}\left|\left\{ v\in\mathbb{Z}^{n}:\,\Lambda^{-1}C_{k,v}-U\subseteq A\right\} \right|.
\end{align*}
Assuming $k\ge-T$ and taking expectation over $U$, we obtain{\allowdisplaybreaks
\begin{align*}
  \E\left[\sum_{l=-\infty}^{k}2^{-nl}\left|D_{l}\left(\Lambda A+U\right)\right|\,\biggl|\,\Lambda\right]
 & =2^{-nT}\int_{[0,2^{T}]^{n}}2^{-nk}\left|\left\{ v\in\mathbb{Z}^{n}:\, C_{k,v}-u\subseteq\Lambda A\right\} \right|du\\
 & =2^{-nT}\int_{[0,2^{T}]^{n}}2^{-nk}\left|\left\{ v\in\mathbb{Z}^{n}:\,2^{-k}\left([0,1]^{n}+v-2^{k}u\right)\subseteq\Lambda A\right\} \right|du\\
 & =\int_{[0,1]^{n}}2^{-nk}\left|\left\{ v\in\mathbb{Z}^{n}:\,2^{-k}\left([0,1]^{n}+v-2^{T+k}u\right)\subseteq\Lambda A\right\} \right|du\\
 & \stackrel{\mathrm{(a)}}{=}\int_{\mathbb{R}^{n}}2^{-nk}\mathbf{1}\left\{ 2^{-k}\left([0,1]^{n}+u\right)\subseteq\Lambda A\right\} du\\
 & =\int_{\mathbb{R}^{n}}\mathbf{1}\left\{ 2^{-k}[0,1]^{n}+u\subseteq\Lambda A\right\} du\\
 & =\mathrm{V}_{n}\left(\Lambda A\ominus2^{-k}[0,1]^{n}\right)\\
 & =\Lambda^{n}\mathrm{V}_{n}\left(A\ominus\Lambda^{-1}2^{-k}[0,1]^{n}\right),
\end{align*}
where $(a)$ follows since $2^{T+k}$ is a non-negative integer. Since $2^{nT}>\mathrm{V}_{n}(2A)$,
$D_{k}(\Lambda A+U)=\emptyset$ for $k<-T$. Hence, we have
\begin{align*}
H(W_{\Lambda A+U}) & =\log\mathrm{V}_{n}(\Lambda A)+\frac{1}{\mathrm{V}_{n}(\Lambda A)}\sum_{k=-T}^{\infty}nk2^{-nk}\left|D_{k}(\Lambda A+U)\right|\\
 & =\log\mathrm{V}_{n}(A)+n\log\Lambda+\frac{n}{\Lambda^{n}\mathrm{V}_{n}(A)}\sum_{k=-T}^{\infty}\left(\mathbf{1}\left\{ k\ge0\right\} \Lambda^{n}\mathrm{V}_{n}(A)-\sum_{l=-\infty}^{k}2^{-nl}\left|D_{l}(\Lambda A+U)\right|\right).
\end{align*}
Taking expectation over $U$, we obtain
\begin{align*}
  \E\left[H(W_{\Lambda A+U})\,\biggl|\,\Lambda\right]
 & =\log\mathrm{V}_{n}(A)+n\log\Lambda+\frac{n}{\Lambda^{n}\mathrm{V}_{n}(A)}\sum_{k=-T}^{\infty}\left(\mathbf{1}\left\{ k\ge0\right\} \Lambda^{n}\mathrm{V}_{n}(A)-\Lambda^{n}\mathrm{V}_{n}\left(A\ominus\Lambda^{-1}2^{-k}[0,1]^{n}\right)\right)\\
 & =\log\mathrm{V}_{n}(A)+n\log\Lambda+n\sum_{k=-T}^{\infty}\left(\mathbf{1}\left\{ k\ge0\right\} -\frac{\mathrm{V}_{n}\left(A\ominus\Lambda^{-1}2^{-k}[0,1]^{n}\right)}{\mathrm{V}_{n}(A)}\right).
\end{align*}
Taking expectation over $\Lambda$, we have
\begin{align*}
  \E\left[H(W_{\Lambda A+U})\right]
 & =\log\mathrm{V}_{n}(A)+\E\left[n\log\Lambda\right]+n\E\left[\sum_{k=-T}^{\infty}\left(\mathbf{1}\left\{ k\ge0\right\} -\frac{\mathrm{V}_{n}\left(A\ominus2^{-(k+\Theta)}[0,1]^{n}\right)}{\mathrm{V}_{n}(A)}\right)\right]\\
 & =\log\mathrm{V}_{n}(A)+\frac{n}{2}+n\left(\int_{-T}^{\infty}\left(\mathbf{1}\left\{ \theta\ge0\right\} -\frac{\mathrm{V}_{n}\left(A\ominus2^{-\theta}[0,1]^{n}\right)}{\mathrm{V}_{n}(A)}\right)d\theta-\frac{1}{2}\right)\\
 & =\log\mathrm{V}_{n}(A)+n\int_{-T}^{\infty}\left(\mathbf{1}\left\{ \theta\ge0\right\} -\frac{\mathrm{V}_{n}\left(A\ominus2^{-\theta}[0,1]^{n}\right)}{\mathrm{V}_{n}(A)}\right)d\theta\\
 & \stackrel{\mathrm{(a)}}{=}\log\mathrm{V}_{n}(A)+n\int_{-\infty}^{\infty}\left(\mathbf{1}\left\{ \theta\ge0\right\} -\frac{\mathrm{V}_{n}\left(A\ominus2^{-\theta}[0,1]^{n}\right)}{\mathrm{V}_{n}(A)}\right)d\theta\\
 & =\log\mathrm{V}_{n}(A)+n\cdot h_{\ominus[0,1]^{n}}(A).
\end{align*}
where $(a)$ follows since $2^{nT}>\mathrm{V}_{n}(2A)$, hence $A\ominus2^{-\theta}[0,1]^{n}=\emptyset$
for $\theta\le-T$.}

\subsection{Proof of Theorem \ref{thm:scale_min} \label{sub:scale_min}}

We first prove the following claim on $H(W_{A})$ involving truncated differential entropy 
\[
H(W_{A})\le n\left(H(\zeta_{1},\ldots,\zeta_{n})+\sum_{i=1}^{n}\zeta_{i}\tilde{h}_{\zeta_{i}}(X_{[1:n]\backslash i})\right)-(n-1)\log\mathrm{V}_{n}(A)+(2+\log e)n,
\]
where 
\[
\zeta_{i}=\int_{\mathbb{R}^{n-1}}\min\left\{ f_{X_{[1:n]\backslash i}}(x_{[1:n]\backslash i})\,,\,\xi\right\} dx_{[1:n]\backslash i}
\]
for a suitable $\xi>0$ such that $\sum\zeta_{i}=1$. By Proposition
\ref{prop:hdy_erosion}, the claim can be proved by bounding $h_{\ominus[0,1]^{n}}(A)$.
Note that by Lemma \ref{lem:ortho_onecon},{\allowdisplaybreaks
\begin{align*}
  h_{\ominus[0,1]^{n}}(A)
 & =\int_{-\infty}^{\infty}\left(\mathbf{1}\left\{ t\ge0\right\} -\frac{\mathrm{V}_{n}\left(A\ominus[0,\,2^{-t}]^{n}\right)}{\mathrm{V}_{n}(A)}\right)dt\\
 & \le\int_{-\infty}^{\infty}\left(\mathbf{1}\left\{ t\ge0\right\} -\frac{1}{\mathrm{V}_{n}(A)}\max\left(0,\,\mathrm{V}_{n}(A)-\sum_{i=1}^{n}\int_{\mathrm{P}_{\backslash i}(A)}\min\left\{ 2^{-t}\,,\,\mathrm{V}_{1}\left(A\cap(\mathrm{span}(\mathrm{e}_{i})+x)\right)\right\} dx_{[1:n]\backslash i}\right)\right)dt\\
 & =\int_{-\infty}^{\infty}\left(\mathbf{1}\left\{ t\ge0\right\} -\max\left(0,\,1-\sum_{i=1}^{n}\int_{\mathrm{P}_{\backslash i}(A)}\min\left\{ \frac{2^{-t}}{\mathrm{V}_{n}(A)}\,,\, f_{X_{[1:n]\backslash i}}(x_{[1:n]\backslash i})\right\} dx_{[1:n]\backslash i}\right)\right)dt\\
 & \stackrel{(a)}{=}\int_{-\infty}^{\infty}\left(-\mathbf{1}\left\{ t<0\right\} +\sum_{i=1}^{n}\int_{\mathrm{P}_{\backslash i}(A)}\min\left\{ \frac{2^{-t}}{\mathrm{V}_{n}(A)},\,\xi,\, f_{X_{[1:n]\backslash i}}(x_{[1:n]\backslash i})\right\} dx_{[1:n]\backslash i}\right)dt\\
 & =\sum_{i=1}^{n}\int_{-\infty}^{\infty}\left(-\mathbf{1}\left\{ t<0\right\} \cdot\zeta_{i}+\int_{\mathrm{P}_{\backslash i}(A)}\min\left\{ \frac{2^{-t}}{\mathrm{V}_{n}(A)},\,\xi,\, f_{X_{[1:n]\backslash i}}(x_{[1:n]\backslash i})\right\} dx_{[1:n]\backslash i}\right)dt\\
 & =\sum_{i=1}^{n}\int_{\mathrm{P}_{\backslash i}(A)}\int_{-\infty}^{\infty}\left(-\mathbf{1}\left\{ t<0\right\} \cdot\min\left\{ \xi,\, f_{X_{[1:n]\backslash i}}(x_{[1:n]\backslash i})\right\} +\min\left\{ \frac{2^{-t}}{\mathrm{V}_{n}(A)},\,\xi,\, f_{X_{[1:n]\backslash i}}(x_{[1:n]\backslash i})\right\} \right)dtdx_{[1:n]\backslash i}\\
 & =\sum_{i=1}^{n}\int_{\mathrm{P}_{\backslash i}(A)}-\min\left\{ \xi,\, f_{X_{[1:n]\backslash i}}(x_{[1:n]\backslash i})\right\} \cdot\left(\log\left(\mathrm{V}_{n}(A)\min\left\{ \xi,\, f_{X_{[1:n]\backslash i}}(x_{[1:n]\backslash i})\right\} \right)-\log e\right)dx_{[1:n]\backslash i}\\
 & =-\log\mathrm{V}_{n}(A)+H(\zeta_{1},\ldots,\zeta_{n})\\
 & \;\;\;\;\;\;+\sum_{i=1}^{n}\zeta_{i}\int_{\mathrm{P}_{\backslash i}(A)}-\zeta_{i}^{-1}\min\left\{ \xi,\, f_{X_{[1:n]\backslash i}}(x_{[1:n]\backslash i})\right\} \cdot\log\left(\zeta_{i}^{-1}\min\left\{ \xi,\, f_{X_{[1:n]\backslash i}}(x_{[1:n]\backslash i})\right\} \right)dx_{[1:n]\backslash i}+\log e\\
 & =-\log\mathrm{V}_{n}(A)+H(\zeta_{1},\ldots,\zeta_{n})+\sum_{i=1}^{n}\zeta_{i}\tilde{h}_{\zeta_{i}}(X_{[1:n]\backslash i})+\log e,
\end{align*}
where $(a)$ follows by the definition of $\xi$. The claim follows.

We proceed to prove Theorem \ref{thm:scale_min}. Let $D=\mathrm{diag}(d_{1},\ldots,d_{n})$.
Assuming $\prod_{i}d_{i}=1$, then by the claim, 
\[
\begin{aligned}H(W_{DA}) & \le n\left(H(\zeta_{1},\ldots,\zeta_{n})+\sum_{i=1}^{n}\zeta_{i}\left(\tilde{h}_{\zeta_{i}}(X_{[1:n]\backslash i})+\sum_{j\neq i}\log d_{j}\right)\right)-(n-1)\log\mathrm{V}_{n}(A)+(2+\log e)n\\
 & \le n\sum_{i=1}^{n}\zeta_{i}\left(\tilde{h}_{\zeta_{i}}(X_{[1:n]\backslash i})-\log d_{i}\right)-(n-1)\log\mathrm{V}_{n}(A)+n\log n+(2+\log e)n,
\end{aligned}
\]
where 
\[
\begin{aligned}\zeta_{i} & =\left(\prod_{j\neq i}d_{j}\right)\int_{\mathbb{R}^{n-1}}\min\left\{ \left(\prod_{j\neq i}d_{j}\right)^{-1}f_{X_{[1:n]\backslash i}}(x_{[1:n]\backslash i})\,,\,\xi\right\} dx_{[1:n]\backslash i}\\
 & =\int_{\mathbb{R}^{n-1}}\min\left\{ f_{X_{[1:n]\backslash i}}(x_{[1:n]\backslash i})\,,\,\xi d_{i}^{-1}\right\} dx_{[1:n]\backslash i}
\end{aligned}
\]
for a suitable $\xi>0$ such that $\sum\zeta_{i}=1$. Let $\alpha_{1},\ldots,\alpha_{n}>0$
such that 
\[
\int_{\mathbb{R}^{n-1}}\min\left\{ f_{X_{[1:n]\backslash i}}(x_{[1:n]\backslash i})\,,\,\alpha_{i}\right\} dx_{[1:n]\backslash i}=\frac{1}{n}.
\]
Set $\xi=\left(\prod_{j}a_{j}\right)^{1/n}$, $d_{i}=\alpha_{i}^{-1}\xi$,
then we have $\zeta_{i}=1/n$, 
\[
H(W_{DA})\le\sum_{i=1}^{n}\tilde{h}_{1/n}(X_{[1:n]\backslash i})-(n-1)\log\mathrm{V}_{n}(A)+n\log n+(2+\log e)n.
\]
}

\subsection{Proof of Lemma \ref{lem:logconcave_cond} \label{sub:logconcave_cond}}

Before proving this lemma, we first prove the following claim on the
volume of a convex set. For any convex set $A\subseteq\mathbb{R}^{n}$
where $0\in A$ and $1\le m\le n$, let $\tilde{A}=\left\{ x_{m+1}^{n}:\,(0^{m},\, x_{m+1}^{n})\in A\right\} $,
then 
\[
\mathrm{V}_{n}(A)\ge {\binom{n}{m}}^{-1} \cdot\mathrm{V}_{n-m}(\tilde{A})\cdot\mathrm{VP}_{[1:m]}(A).
\]
Now we prove the claim. Denote the section of $A$ as {\allowdisplaybreaks
\[
S_{A}(\tilde{x}_{m+1}^{n})=\left\{ x^{m}:\,(x^{m},\,\tilde{x}_{m+1}^{n})\in A\right\} \subseteq\mathbb{R}^{m}.
\]
Note that {\allowdisplaybreaks
\begin{align*}
\mathrm{V}_{n}(A) & =\int_{S^{m-1}}\int_{0}^{\infty}\left(\int_{\left\{ \tilde{x}_{m+1}^{n} :\, (rx^{m},\,\tilde{x}_{m+1}^{n})\in A\right\} }d\tilde{x}_{m+1}^{n}\right)r^{m-1}drdx^{m}\\
 & =\int_{S^{m-1}}\int_{\left\{ (r,\,\tilde{x}_{m+1}^{n}):\,r\ge0,\,(rx^{m},\,\tilde{x}_{m+1}^{n})\in A\right\} }r^{m-1}d(r,\,\tilde{x}_{m+1}^{n})dx^{m}.
\end{align*}}
Consider the set 
\[
S_{\mathrm{rad},A}(x^{m})=\left\{ (r,\,\tilde{x}_{m+1}^{n}):\, r\ge0,\,(rx^{m},\,\tilde{x}_{m+1}^{n})\in A\right\} .
\]
It is the intersection of $A$ and a half-space, and hence it is convex.
By definition of radial function and projection, there exists $\hat{x}_{m+1}^{n}(x^{n})$
such that 
\[
(\rho_{\mathrm{P}_{[1:m]}(A)}(x^{m}),\,\hat{x}_{m+1}^{n}(x^{n}))\in S_{\mathrm{rad},A}(x^{m}).
\]
Also by definition of $\tilde{A}$, 
\[
\left\{ 0\right\} \times\tilde{A}\subseteq S_{\mathrm{rad},A}(x^{m}).
\]
Hence the convex hull of $\bigl\{(\rho_{\mathrm{P}_{[1:m]}(A)}(x^{m}),\,\hat{x}_{m+1}^{n}(x^{n}))\bigr\}\cup\left(\left\{ 0\right\} \times\tilde{A}\right)$
is a subset of $S_{\mathrm{rad},A}(x^{m})$. The convex hull can
be expressed as 
\[
\left\{ (r,\,\tilde{x}_{m+1}^{n}):\,0\le r\le\rho_{\mathrm{P}_{[1:m]}(A)}(x^{m}),\,\tilde{x}_{m+1}^{n}\in\left(1-r\rho_{\mathrm{P}_{[1:m]}(A)}^{-1}(x^{m})\right)\tilde{A}+r\rho_{\mathrm{P}_{[1:m]}(A)}^{-1}(x^{m})\cdot\hat{x}_{m+1}^{n}(x^{n})\right\} .
\]
Therefore, 
\[
\begin{aligned}\mathrm{V}_{n}(A) & =\int_{S^{m-1}}\int_{S_{\mathrm{rad},A}(x^{m})}r^{m-1}d(r,\,\tilde{x}_{m+1}^{n})dx^{m}\\
 & \ge\int_{S^{m-1}}\int_{0}^{\rho_{\mathrm{P}_{[1:m]}(A)}(x^{m})}\int_{\left(1-r\rho_{\mathrm{P}_{[1:m]}(A)}^{-1}(x^{m})\right)\tilde{A}+r\rho_{\mathrm{P}_{[1:m]}(A)}^{-1}(x^{m})\cdot\hat{x}_{m+1}^{n}(x^{n})}d\tilde{x}_{m+1}^{n}\cdot r^{m-1}drdx^{m}\\
 & =\int_{S^{m-1}}\int_{0}^{\rho_{\mathrm{P}_{[1:m]}(A)}(x^{m})}\mathrm{V}_{n-m}(\tilde{A})\left(1-r\rho_{\mathrm{P}_{[1:m]}(A)}^{-1}(x^{m})\right)^{n-m}r^{m-1}drdx^{m}\\
 & =\int_{S^{m-1}}\rho_{\mathrm{P}_{[1:m]}(A)}^{m}(x^{m})\mathrm{V}_{n-m}(\tilde{A})\cdot\mathrm{B}(m,\, n-m+1)dx^{m}\\
 & =m\cdot\mathrm{B}(m,\, n-m+1)\cdot\mathrm{V}_{n-m}(\tilde{A})\cdot\mathrm{VP}_{[1:m]}(A),
\end{aligned}
\]
where 
\[
\mathrm{B}(\alpha,\beta)=\int_{0}^{1}t^{\alpha-1}(1-t)^{\beta-1}dt
\]
is the beta function.
The claim follows.

We proceed to prove Lemma \ref{lem:logconcave_cond}. To prove the
lower bound, consider
\[
\begin{aligned}h(X_{m+1}^{n}\,|\, X^{m}) & =\int_{\mathbb{R}^{m}}f_{X^{m}}(x^{m})h(X_{m+1}^{n}\,|\, X^{m}=x^{m})dx^{m}\\
 & \ge\int_{\mathbb{R}^{m}}f_{X^{m}}(x^{m})\cdot-\log\sup_{\tilde{x}_{m+1}^{n}\in\mathbb{R}^{n-m}}f(\tilde{x}_{m+1}^{n}\,|\, x^{m})dx^{m}\\
 & \ge-\log\int_{\mathbb{R}^{m}}f_{X^{m}}(x^{m})\sup_{\tilde{x}_{m+1}^{n}\in\mathbb{R}^{n-m}}f(\tilde{x}_{m+1}^{n}\,|\, x^{m})dx^{m}\\
 & =-\log\int_{\mathbb{R}^{m}}\sup_{\tilde{x}_{m+1}^{n}\in\mathbb{R}^{n-m}}f(x^{m},\,\tilde{x}_{m+1}^{n})dx^{m}.
\end{aligned}
\]
Now we prove the upper bound. By Lemma \ref{lem:logconcave_sup},
\begin{equation}
h(X^{n})\le-\log\Bigl(\sup_{x^{n}\in\mathbb{R}^{n}}f(x^{n})\Bigr)+n\log e,\label{eq:logconcave_cond_hxn}
\end{equation}
\begin{equation}
h(X^{m})\ge-\log\Bigl(\sup_{x^{m}\in\mathbb{R}^{m}}f_{X^{m}}(x^{m})\Bigr).\label{eq:logconcave_cond_hxm}
\end{equation}
Without loss of generality, assume that $\sup_{x^{m}\in\mathbb{R}^{m}}f_{X^{m}}(x^{m})$
is attained at $x^{m}=0$ and $\sup_{x_{m+1}^{n}}f(0^{m},x_{m+1}^{n})$
is attained at $x_{m+1}^{n}=0$. Denote the super level set 
\[
L_{z}^{+}(f)=\left\{ x^{n}:\, f(x^{n})\ge z\right\} .
\]
Since $f$ is log-concave, $L_{z}^{+}(f)$ is convex. Define 
\[
\tilde{L}_{z}^{+}(f)=\left\{ x_{m+1}^{n}:\,(0^{m},\, x_{m+1}^{n})\in L_{z}^{+}(f)\right\} .
\]
By the claim we proved earlier, {\allowdisplaybreaks
\begin{align*}\int_{\mathbb{R}^{n}}f(x^{n}) & =\int_{0}^{\infty}\mathrm{V}_{n}\left(L_{z}^{+}(f)\right)dz\\
 & \ge\int_{\left\{ z:\,0\in L_{z}^{+}(f)\right\} }\mathrm{V}_{n}\left(L_{z}^{+}(f)\right)dz\\
 & \ge\int_{\left\{ z:\,0\in L_{z}^{+}(f)\right\} } {\binom{n}{m}}^{-1}\cdot\mathrm{V}_{n-m}(\tilde{L}_{z}^{+}(f))\cdot\mathrm{VP}_{[1:m]}(L_{z}^{+}(f))dz\\
 & ={\binom{n}{m}}^{-1}\cdot\int_{0}^{f(0)}\mathrm{V}_{n-m}(\tilde{L}_{z}^{+}(f))\cdot\mathrm{VP}_{[1:m]}(L_{z}^{+}(f))dz\\
 & \stackrel{\mathrm{(a)}}{\ge}{\binom{n}{m}}^{-1}\cdot\left(\int_{0}^{f(0)}\mathrm{V}_{n-m}(\tilde{L}_{z}^{+}(f))dz\right)\cdot\left(\frac{1}{f(0)}\int_{0}^{f(0)}\mathrm{VP}_{[1:m]}(L_{z}^{+}(f))dz\right)\\
 & \stackrel{\mathrm{(b)}}{\ge}{\binom{n}{m}}^{-1}\cdot f_{X^{m}}(0)\cdot\left(\frac{1}{\sup_{x}f(x)}\int_{0}^{\sup_{x}f(x)}\mathrm{VP}_{[1:m]}(L_{z}^{+}(f))dz\right)\\
 & ={\binom{n}{m}}^{-1}\cdot f_{X^{m}}(0)\cdot\frac{1}{\sup_{x}f(x)}\int_{\mathbb{R}^{m}}\sup_{\tilde{x}_{m+1}^{n}\in\mathbb{R}^{n-m}}f(x^{m},\,\tilde{x}_{m+1}^{n})dx^{m}\\
 & \stackrel{\mathrm{(c)}}{\ge}{\binom{n}{m}}^{-1}\cdot2^{-h(X^{m})}\cdot2^{h(X^{n})}e^{-n}\int_{\mathbb{R}^{m}}\sup_{\tilde{x}_{m+1}^{n}\in\mathbb{R}^{n-m}}f(x^{m},\,\tilde{x}_{m+1}^{n})dx^{m}.
\end{align*}}
where $(a)$ is due to Chebyshev's sum inequality, since both $\mathrm{V}_{n-m}(\tilde{L}_{z}^{+}(f))$
and $\mathrm{VP}_{[1:m]}(L_{z}^{+}(f))$ are non-increasing in $z$,
$(b)$ is due to $\int_{0}^{f(0)}\mathrm{V}_{n-m}(\tilde{L}_{z}^{+}(f))dz=\int_{\mathbb{R}^{n-m}}f(0^{m},\,\tilde{x}_{m+1}^{n})d\tilde{x}_{m+1}^{n}=f_{X^{m}}(0)$
since $\sup_{\tilde{x}_{m+1}^{n}}f(0^{m},\tilde{x}_{m+1}^{n})=f(0)$,
and $\mathrm{VP}_{[1:m]}(L_{z}^{+}(f))\, dz$ is non-increasing in $z$,
and $(c)$ is due to (\ref{eq:logconcave_cond_hxn}) and (\ref{eq:logconcave_cond_hxm}).
The result follows from $\int_{\mathbb{R}^{n}}f(x^{n})=1$.}

\subsection{Proof of Lemma \ref{lem:logconcave_trunc} \label{sub:logconcave_trunc}}

As in the definition of $\tilde{h}_{\zeta}$, let $\xi>0$ such that
\[
\int_{\mathbb{R}^{n}}\min\left\{ \xi,\, f(x)\right\} dx=\zeta.
\]
Without loss of generality, assume $\sup_{x\in\mathbb{R}^{n}}f(x)=f(0)$
and let $\alpha=f(0)$. Let 
\[
A=\left\{ x\in\mathbb{R}^{n}:\, f(x)\ge\xi\right\} .
\]
By log-concavity of $f$, we know $A$ is convex, and we have 
\[
\mathrm{V}_{n}(A)=\frac{1}{n}\cdot\int_{S^{n-1}}\rho_{A}^{n}(x)dx,
\]
where $S^{n-1}=\left\{ x\in\mathbb{R}^{n}\,:\,\left\Vert x\right\Vert _{2}=1\right\} $
is the unit sphere. For $x\in A$, by definition of $\rho_{A}$, we
know $x/(\rho_{A}^{-1}(x)+\epsilon)\in A$ for any $\epsilon>0$,
\[
\begin{aligned}f(x) & =f\left(\left(1-\rho_{A}^{-1}(x)-\epsilon\right)\cdot0+\left(\rho_{A}^{-1}(x)+\epsilon\right)\cdot\frac{x}{\rho_{A}^{-1}(x)+\epsilon}\right)\\
 & \ge\left(f(0)\right)^{1-\rho_{A}^{-1}(x)-\epsilon}\cdot\left(f\left(\frac{x}{\rho_{A}^{-1}(x)+\epsilon}\right)\right)^{\rho_{A}^{-1}(x)+\epsilon}\\
 & \ge\alpha^{1-\rho_{A}^{-1}(x)-\epsilon}\xi^{\rho_{A}^{-1}(x)+\epsilon}.
\end{aligned}
\]
Therefore 
\[
f(x)\ge\alpha^{1-\rho_{A}^{-1}(x)}\xi^{\rho_{A}^{-1}(x)}.
\]
Hence 
\[
\begin{aligned}\int_{A}f(x)dx & =\int_{S^{n-1}}\int_{0}^{\rho_{A}(x)}f(rx)\cdot r^{n-1}drdx\\
 & \ge\int_{S^{n-1}}\int_{0}^{\rho_{A}(x)}\alpha^{1-\rho_{A}^{-1}(rx)}\xi^{\rho_{A}^{-1}(rx)}r^{n-1}drdx\\
 & =\int_{S^{n-1}}\int_{0}^{\rho_{A}(x)}\alpha^{1-r\rho_{A}^{-1}(x)}\xi^{r\rho_{A}^{-1}(x)}r^{n-1}drdx\\
 & =\int_{S^{n-1}}\rho_{A}^{n}(x)\int_{0}^{1}\alpha^{1-r}\xi^{r}r^{n-1}drdx\\
 & =\int_{S^{n-1}}\rho_{A}^{n}(x)\left(\alpha\left(-\log(\xi/\alpha)\right)^{-n}\left(\Gamma(n)-\Gamma(n,\,-\log(\xi/\alpha))\right)\right)dx\\
 & =n\alpha\left(-\log(\xi/\alpha)\right)^{-n}\left(\Gamma(n)-\Gamma(n,\,-\log(\xi/\alpha))\right)\mathrm{V}_{n}(A),
\end{aligned}
\]
where $\Gamma(n,z)=\int_{z}^{\infty}t^{n-1}e^{-t}dt$ is the incomplete
gamma function, and $\Gamma(n)=\Gamma(n,0)$ is the gamma function.

On the other hand, for $x\notin A$, then for any $\epsilon>0$, we
have $x/(\rho_{A}^{-1}(x)-\epsilon)\notin A$, 
\[
\begin{aligned}\xi\ge f\left(\frac{x}{\rho_{A}^{-1}(x)-\epsilon}\right) & =f\left(\left(1-\frac{1}{\rho_{A}^{-1}(x)-\epsilon}\right)\cdot0+\frac{1}{\rho_{A}^{-1}(x)-\epsilon}\cdot x\right)\\
 & \ge\alpha^{1-1/(\rho_{A}^{-1}(x)-\epsilon)}\cdot\left(f(x)\right)^{1/(\rho_{A}^{-1}(x)-\epsilon)},
\end{aligned}
\]
Therefore 
\[
f(x)\le\alpha^{1-\rho_{A}^{-1}(x)}\xi^{\rho_{A}^{-1}(x)}.
\]
Hence 
\[
\begin{aligned}\int_{\mathbb{R}^{n}\backslash A}f(x)dx & =\int_{S^{n-1}}\int_{\rho_{A}(x)}^{\infty}f(rx)\cdot r^{n-1}drdx\\
 & \le\int_{S^{n-1}}\int_{\rho_{A}(x)}^{\infty}\alpha^{1-\rho_{A}^{-1}(rx)}\xi^{\rho_{A}^{-1}(rx)}r^{n-1}drdx\\
 & =\int_{S^{n-1}}\int_{\rho_{A}(x)}^{\infty}\alpha^{1-r\rho_{A}^{-1}(x)}\xi^{r\rho_{A}^{-1}(x)}r^{n-1}drdx\\
 & =\int_{S^{n-1}}\rho_{A}^{n}(x)\int_{1}^{\infty}\alpha^{1-r}\xi^{r}r^{n-1}drdx\\
 & =\int_{S^{n-1}}\rho_{A}^{n}(x)\left(\alpha\left(-\log(\xi/\alpha)\right)^{-n}\Gamma(n,\,-\log(\xi/\alpha))\right)dx\\
 & =n\alpha\left(-\log(\xi/\alpha)\right)^{-n}\Gamma(n,\,-\log(\xi/\alpha))\mathrm{V}_{n}(A).
\end{aligned}
\]
Recall that $\int_{\mathbb{R}^{n}}\min\left\{ \xi,\, f(x)\right\} dx=\zeta$,
\[
\begin{aligned}\zeta & =\int_{\mathbb{R}^{n}}\min\left\{ \xi,\, f(x)\right\} dx\\
 & =\int_{\mathbb{R}^{n}\backslash A}f(x)dx+\xi\mathrm{V}_{n}(A)\\
 & \le\left(n\alpha\left(-\log(\xi/\alpha)\right)^{-n}\Gamma(n,\,-\log(\xi/\alpha))+\xi\right)\mathrm{V}_{n}(A).
\end{aligned}
\]
Also 
\[
\begin{aligned}\zeta & =\int_{\mathbb{R}^{n}}\min\left\{ \xi,\, f(x)\right\} dx\\
 & =1-\int_{A}f(x)dx+\xi\mathrm{V}_{n}(A)\\
 & \le1-\left(n\alpha\left(-\log(\xi/\alpha)\right)^{-n}\left(\Gamma(n)-\Gamma(n,\,-\log(\xi/\alpha))\right)-\xi\right)\mathrm{V}_{n}(A).
\end{aligned}
\]
Let $\nu=-\log(\xi/\alpha)$. Since $\zeta\le ac$ and $\zeta\le1-bc$
implies $\zeta\le a/(a+b)$, we have 
\[
\begin{aligned}\zeta & \le\frac{n\alpha\nu^{-n}\Gamma(n,\nu)+\xi}{\left(n\alpha\nu^{-n}\Gamma(n,\nu)+\xi\right)+\left(n\alpha\nu^{-n}\left(\Gamma(n)-\Gamma(n,\nu)\right)-\xi\right)}\\
 & =\frac{n\Gamma(n,\nu)+e^{-\nu}\nu^{n}}{n\Gamma(n)}\\
 & =\frac{\Gamma(n+1,\,\nu)}{\Gamma(n+1)}.
\end{aligned}
\]
By Lemma \ref{lem:logconcave_sup}, $h(X^{n})\ge-\log\alpha$.
Recall that $\tilde{h}_{\zeta}(X^{n})$ is the entropy of the
pdf $\tilde{f}(x)=\zeta^{-1}\min\left\{ \xi,\, f(x)\right\} $, which
is also log-concave. Hence by Lemma \ref{lem:logconcave_sup}, $\tilde{h}_{\zeta}(X^{n})\le-\log(\zeta^{-1}\xi)+n\log e$.
As a result, 
\[
\begin{aligned}\tilde{h}_{\zeta}(X^{n})-h(X^{n}) & \le-\log\Bigl(\zeta^{-1}\xi/\alpha\Bigr)+n\log e\\
 & \le\log\zeta+\nu+n\log e.
\end{aligned}
\]

To prove the second bound, assume that $\zeta\ge e^{-(e-2)n}$ and $\nu>en$.
We use the bound 
\[
\Gamma(a,z)<Bz^{a-1}e^{-z}
\]
 for $a>1$, $B>1$, $z>B(a-1)/(B-1)$ due to \cite{natalini2000inequalities}.

Substituting $a=n+1$, $z=en$, and $B=e$, we have $\Gamma(n+1,\nu)<\Gamma(n+1,en)<e(en)^{n}e^{-en}$.
We also know that $\Gamma(n+1)\ge n^{n}e^{-(n-1)}$, hence 
\[
\begin{aligned}\frac{\Gamma(n+1,\,\nu)}{\Gamma(n+1)} & <\frac{e(en)^{n}e^{-en}}{n^{n}e^{-(n-1)}}\\
 & =e^{-(e-2)n},
\end{aligned}
\]
which leads to a contradiction and $\nu\le en$. The result follows.

 \bibliographystyle{IEEEtran}
\bibliography{ref,nit}

\end{document}